\documentclass[superscriptaddress,showpacs,aps,twocolumn,amsmath,amsfonts,floatfix,xcolor=dvipsnames]{revtex4-1}
 
\usepackage{graphicx}           
\usepackage{color}
\usepackage[dvipsnames]{xcolor}
\usepackage{amsmath}
\usepackage
	[colorlinks=true,
	urlcolor=blue,
	linkcolor=blue,
    citecolor=blue]
    {hyperref}
\usepackage{revsymb4-1}
\usepackage{rotating}
\usepackage{float}
\usepackage{setspace}
\usepackage{multirow}
\usepackage{wasysym}
\usepackage{bm}
\usepackage{fixmath}
\usepackage{placeins}
\usepackage{amsbsy}
\usepackage{pifont}
\usepackage{hhline}
\usepackage{tabularx}
\usepackage{MnSymbol}
\usepackage{ulem}
\usepackage{nicefrac}
\begin{document}

\title{
Anisotropy crossover in the frustrated Hubbard model on four-chain cylinders 
}

\author{
G.\ Ehlers
}
\affiliation{
Fachbereich Physik, Philipps-Universit\"{a}t Marburg, 35032 Marburg, Germany
}
\author{
B.\ Lenz
}
\affiliation{
Centre de Physique Th\'{e}orique, Ecole Polytechnique, CNRS, 
Universit\'{e} Paris Saclay, 91228 Palaiseau, France
}
\affiliation{
Institut f\"{u}r Theoretische Physik, 
Universit\"{a}t G\"{o}ttingen, 37077 G\"{o}ttingen, Germany
}
\author{
S.\ R.\ Manmana
}
\affiliation{
Institut f\"{u}r Theoretische Physik, 
Universit\"{a}t G\"{o}ttingen, 37077 G\"{o}ttingen, Germany
}
\author{
R.\ M.\ Noack
}
\affiliation{
Fachbereich Physik, Philipps-Universit\"{a}t Marburg, 35032 Marburg, Germany
}


\onecolumngrid

\date{\today}

\begin{abstract}

Motivated by dimensional crossover in layered organic ${\kappa}$ salts,
we determine the phase diagram of a system 
of four periodically coupled Hubbard chains with frustration at half filling
as a function of the interchain hopping ${t_{\perp}/t}$
and interaction strength ${U/t}$
at a fixed ratio of frustration and interchain hopping ${t'/t_{\perp}=-0.5}$.
We cover the range from the one-dimensional limit of uncoupled
chains (${t_{\perp}/t=0.0}$) to the isotropic model (${t_{\perp}/t=1.0}$).
For strong ${U/t}$, we find an antiferromagnetic insulator;
in the weak-to-moderate-interaction regime, 
the phase diagram features quasi-one-dimensional antiferromagnetic behavior,
an incommensurate spin-density wave, 
and a metallic phase as ${t_{\perp}/t}$ is increased.
We characterize the phases through their magnetic ordering,
dielectric response, and dominant static correlations.
Our analysis is based primarily on a variant of the 
density-matrix renormalization-group algorithm  
based on an efficient hybrid--real-momentum-space formulation, 
in which we can treat relatively large lattices albeit of a limited width.
This is complemented by a variational cluster approximation study
with a cluster geometry corresponding to the cylindrical lattice 
allowing us to directly compare the two methods for this geometry.
As an outlook, we make contact with work studying dimensional crossover 
in the full two-dimensional system.

\end{abstract}


\maketitle

\section{Introduction}

How dimensional crossover between one-dimensional and two-dimensional
behavior takes place in antiferromagnetic spin systems and in
fermionic systems with Hubbard interactions is a question that has long been of
interest~\cite{bourbonnais1991,bourbonnais1995,affleck1996renormalization,
clarke1997}, 
especially in the context of ladder and anisotropic cuprate 
systems~\cite{Dagotto1996,Rice1996,Nagata1998,Kojima2001} 
and in Bechgaard salts~\cite{giamarchi2004,Vescoli1998}.
Recent experimental investigations of the universality class 
of the Mott transition in layered organic charge-transfer 
salts~\cite{kagawa2005,kagawa2009,furukawa2015,abdeljawad2015} 
have reopened the question of dimensional crossover, 
but with several new ingredients.
The half filled systems are quasi-two-dimensional, anisotropic, and frustrated.
The relative strength of these ingredients as well as 
of the relative interaction strength can be tuned experimentally by 
changing the molecular composition of the salts, 
by applying physical pressure, and by carrying out chemical
substitution.
In particular, a Mott metal-insulator transition (MIT) takes place
which can be analyzed very precisely by measuring the conductivity
as a function of temperature and ambient pressure~\cite{kanoda2011}.

On the theoretical side, a relatively simple model that is thought to capture
the essential features of dimensional crossover is the
two-dimensional, anisotropic, frustrated Hubbard model at half filling.
Note that for the layered organic $\kappa$ salts, effective
models have been postulated that are based on the anisotropic triangular
lattice~\cite{kandpal2009,nakamura2012}, whereas, for the Bechgaard
and Fabre salts, models based on the anisotropic  square lattice
have been postulated~\cite{jacko2013,brown2015}.
Let us nevertheless first consider what is known about the
isotropic square-lattice case with isotropic frustration. 
Geometrical frustration can most easily be introduced 
by taking the single-band Hubbard model on a
square lattice geometry and adding next-nearest neighbor hopping terms $t'$ 
in either one (equivalent to the anisotropic triangular lattice) 
or in both diagonal directions.
Phase diagrams for both cases have been obtained 
using the cluster dynamical mean-field theory (CDMFT)~\cite{biroli2001CDMFT1}
in Ref.~\cite{tremblay2006frustratedHubbard}, 
the variational cluster approximation (VCA)~\cite{dahnken2003VCA} 
in Refs.~\cite{tremblay2008frustratedHubbard,yamada2013frustratedHubbard}, 
the path-integral renormalization group
(PIRG)~\cite{imada2001PathIntegralRenormalizationGroup} 
in Refs.~\cite{imada2002frustratedHubbard,imada2006frustratedHubbard}, 
the density-matrix embedding theory (DMET)~\cite{chan2012DMET} 
in Ref.~\cite{chan2016HubbardPhaseDiagram}, 
and the determinant quantum Monte Carlo method
(DQMC)~\cite{scalapino1981DQMC} in
Ref.~\cite{tianxing2016frustratedHubbard}.
Mott-insulating, metallic, antiferromagnetic, 
and superconducting phases appear depending on the level of frustration 
and the interaction strength.
The antiferromagnetic phase further subdivides into regions 
of different magnetic ordering characterized by distinct ordering wave
vectors~\cite{imada2006frustratedHubbard,tremblay2008frustratedHubbard,
yamada2013frustratedHubbard}.
Despite the enormous efforts made, 
completely conclusive phase diagrams have yet to be found, 
and the results obtained by the different methods often differ in the details.
Open questions include the exact determination 
of the various magnetic orderings 
and the existence and nature of an intermediate phase 
between the insulating strong-coupling 
and the conducting weak-coupling regimes of the phase diagrams.

We now return to {\it dimensional crossover}, i.e., consider the
effect of anisotropy in the couplings in the spatial directions.
Here we want to make a clear distinction between 
dimensionality and frustration, i.e., be able to treat 
the two effects separately.
Thus, in the following, we treat the square lattice 
geometry with next-nearest-neighbor hopping terms of equal strength in
both diagonal directions.
By tuning the interchain hopping $t_{\perp}$
the crossover between uncoupled chains over
the ``quasi-one-dimensional'' case of weakly coupled chains to
the isotropic two-dimensional model can be
investigated~\cite{assaad2012dimensionalCrossoverFrustratedHubbard,
marcin2013dimensionalCrossoverHubbard,pollet2015DimCross,
lenz2016dimensionalCrossoverFrustratedHubbard}.
In the presence of frustration, the most straightforward parametrization 
is to keep the ratio between the value of the frustration, $t'$, 
and $t_\perp$ fixed. 

A direct motivation for the present work is
Ref.~\cite{lenz2016dimensionalCrossoverFrustratedHubbard}, 
in which quantum critical behavior was identified  
using quantum cluster techniques at zero and finite temperature. 
In particular, the paramagnetic Mott transition was found to change 
from a discontinuous transition at $T_c>0$ at large interchain coupling 
to a continuous transition at zero temperature for small coupling strengths.
In other words, the interchain hopping strength serves as a tuning parameter 
to realize a quantum phase transition when going below a critical value 
$t_{\perp\,c}$.
This is an interesting starting point for further investigations 
of quantum critical behavior,
which is seen in the layered organic charge-transfer salts,
in particular, in the temperature-dependent behavior.

The study of Ref.~\cite{lenz2016dimensionalCrossoverFrustratedHubbard}, however,
has a significant limitation in that only the paramagnetic case was treated,
i.e., the possibility of magnetic ordering was explicitly excluded.
	
The goal of the present study is to go beyond this restriction by utilizing 
an unbiased numerical method, 
the density-matrix renormalization-group (DMRG), 
and by applying a variation of the VCA 
that allows for antiferromagnetic ordering.
Since the variation of the DMRG that we use, 
which works in a hybrid of real and momentum space, 
most efficiently treats lattices with cylindrical geometry, 
we focus on them in both methods.
Our study thus  addresses the effect of antiferromagnetic fluctuations
and benchmarks the VCA against the numerically exact hybrid-space DMRG results. 
Note that our aim here is not to specifically model the $\kappa$ salts;
to do this one would have to treat an anisotropic triangular model. 
Here we investigate the effect of dimensionality systematically, 
which is believed to be---together with frustration---one of the driving
mechanisms of the crossover in these systems.
Due to the limitations of the entropy area
law~\cite{eisert2010areaLaw},
it is exponentially difficult to  make the cylinders wider.
Thus, we generically consider cylinders of width 4 in the following, 
but also present results for cylinders of width 5 
for selected values of the parameters.
In particular, 
we study the ground-state properties as a function of the interaction strength
and the anisotropy, 
fixing the frustrating hopping element ${t^\prime= -0.5 \, t_\perp}$, 
where $t_\perp$ is the interchain hopping.

The low-energy properties of systems of coupled Hubbard chains
can be treated in weak coupling using a field-theoretical treatment
based on the renormalization group and
bosonization~\cite{balents1996WeakChains,balents1997WeakNChains,solyom2006ladders,arrigoni1996phase}.
While Ref.~\cite{balents1997WeakNChains} does work out a detailed
weak-coupling ground-state phase diagram for the four-chain case,
it does not explicitly treat the half filled case, 
in which umklapp processes are relevant, 
and it does not include next-nearest-neighbor hopping, 
which is relevant for us here, in the band structure;
therefore the results are of limited usefulness here.
However,  we will discuss relevant features of the four-chain band structure
with next-nearest-neighbor hopping in the following, 
with a view to making contact with weak-coupling treatments.

The paper is organized as follows:
the model is introduced in Sec.~\ref{sec:model},
and details concerning the DMRG and VCA algorithms used are described in
Secs.~\ref{sec:DMRG} and~\ref{sec:VCA}, respectively.
The results are presented in Sec.~\ref{sec:results},
which is subdivided into Sec.~\ref{sec:dmrg_results}, DMRG results, and 
Sec.~\ref{sec:vca_results}, VCA results. 
Section~\ref{sec:discussion} contains a detailed discussion of the
results and their implications and
Sec.~\ref{sec:conclusion} concludes.

\section{Anisotropic Frustrated Hubbard model}
\label{sec:model}

\begin{figure}	
	\includegraphics[width=8.6cm]{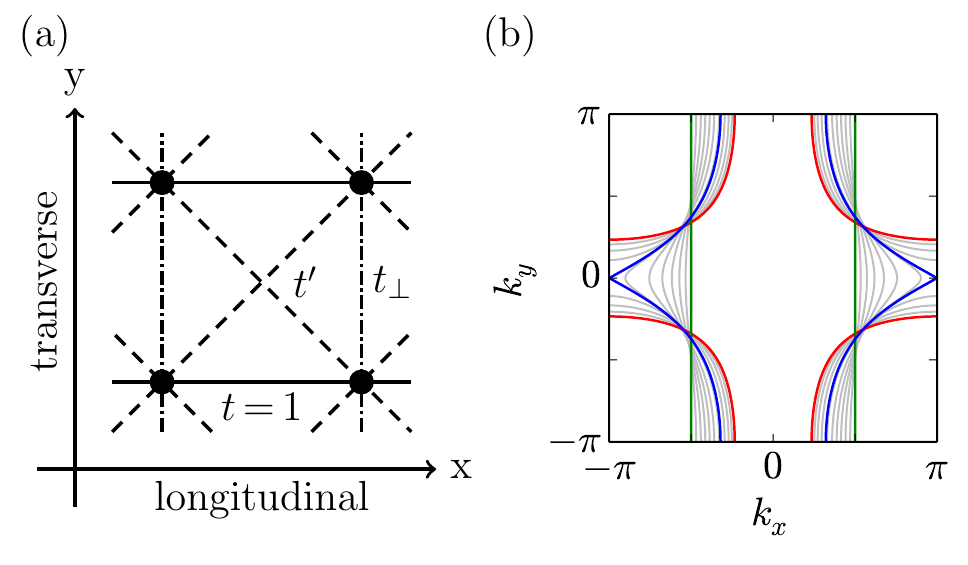}
	\caption{
		(Color online) 
		(a) Depiction of the anisotropic Hubbard model on a square lattice 
		with frustration, indicating the longitudinal hopping $t$,
        which we set to 1, transverse hopping $t_{\perp}$, 
        and diagonal hopping $t^{\prime}$.
		(b) Fermi surface of the noninteracting model at half filling with
		${0\leq t_{\perp} \leq 1}$, and ${t'/t_{\perp}=-0.5}$.
		The green, blue, and red lines indicate 
		${t_{\perp} = 0}$, ${t_{\perp} \approx 0.622}$, 
		and ${t_{\perp} = 1}$, respectively.
	}
	\label{fig:hubbard}
\end{figure}
We treat the anisotropic Hubbard model with frustration, 
\begin{align}
	{\rm H} = &
	-t \sum_{\langle{\bf r},{\bf r}' \rangle_{\parallel}\, \sigma } 
	c^\dagger_{{\bf r} \, \sigma} \, 
	c^{\vphantom{\dagger}}_{{\bf r}' \, \sigma}
	-t_{\perp} \sum_{\langle{\bf r},{\bf r}' \rangle_{\perp}\, \sigma }  
	c^\dagger_{{\bf r} \, \sigma} \, 
	c^{\vphantom{\dagger}}_{{\bf r}' \, \sigma}  \nonumber 
	\\ &
	 -t^\prime \sum_{\llangle{\bf r},{\bf r}' \rrangle \, \sigma } 
	 c^\dagger_{{\bf r} \, \sigma} \, 
	 c^{\vphantom{\dagger}}_{{\bf r}' \, \sigma}
	 + U \sum_{\bf r} 
	 n^{\vphantom{\dagger}}_{{\bf r} \, \uparrow} \,
	 n^{\vphantom{\dagger}}_{{\bf r}\,\downarrow} \,
	 \label{eqn:hubbard} \, ,
\end{align}
where 
	${c^{\dagger}_{{\bf r}\,\sigma}}$, 
	${c^{\hphantom{\dagger}}_{{\bf r}\,\sigma}}$, 
and
	${n^{\hphantom{\dagger}}_{{\bf r}\,\sigma}}=
	{c^{\dagger}_{{\bf r}\,\sigma}} \, 
	{c^{\hphantom{\dagger}}_{{\bf r}\,\sigma}}$ 
are the creation, annihilation, and density operators for lattice site 
	${{\bf r}=(x,y)}$ 
with spin 
	${\sigma \in \left\{\uparrow\,,\,\downarrow\right\}}$.
As depicted in Fig.~\ref{fig:hubbard}(a), 
the amplitudes $t$, ${t_{\perp}}$, and ${t'}$ 
are for hopping in the longitudinal (i.e., intrachain), 
transverse (i.e., interchain), and diagonal directions;
	${\langle{\bf r},{\bf r}' \rangle_{\parallel}}$ 
and 
	${\langle{\bf r},{\bf r}' \rangle_{\perp}}$
denote nearest-neighbor pairs of sites in the longitudinal 
and transverse directions, respectively, 
whereas ${\llangle{\bf r},{\bf r}' \rrangle}$ denotes
next-nearest-neighbor pairs, here along the diagonals.
The dispersion relation for the infinite lattice 
then has the form
\begin{align}
	\varepsilon({\bf k}) & = 
	- 2 t \, \cos k_x 
	- 2 t_{\perp} \, \cos k_y 
	- 4 t^{\prime} \, \cos k_x \, \cos k_y \,,
	\label{eqn:dispersion}
\end{align}
leading to the noninteracting Fermi surface depicted 
for varying $t_\perp$ in Fig.~\ref{fig:hubbard}(b).
We study the model on lattices
with periodic boundary conditions in the transverse direction and open boundary
conditions in the longitudinal directions, i.e., on a cylinder geometry.
Taking the lattice spacing to be unity, 
we denote the length in the longitudinal direction ${L_x}$, 
the width (in the transverse direction) ${L_y}$,
and the number of sites $N = L_x \, L_y$.

\begin{figure}	
	\includegraphics[width=8.6cm]{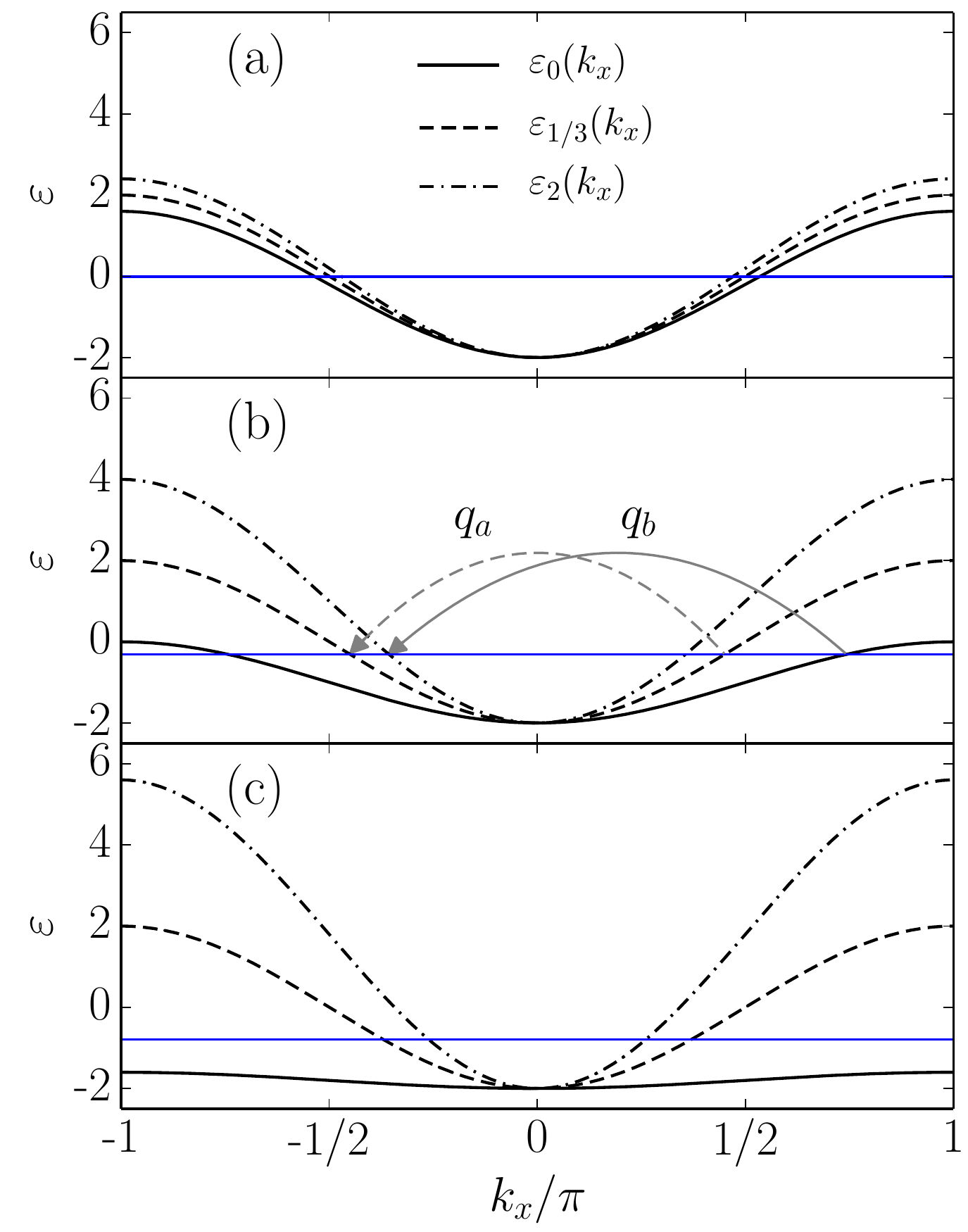}
	\caption{
		(Color online) 
		Band structure for four coupled Hubbard chains
		with frustration ${t'/t_{\perp}=-0.5}$
		and interchain hopping (a) ${t_{\perp}=0.1}$,
		(b) ${t_{\perp}=0.5}$, and (c) ${t_{\perp}=0.9}$.
		The designation ${\varepsilon_j(k_x)}$ refers to a band with
		transverse momentum ${k_y=j\,\pi/2}$. 
		The blue lines indicate the corresponding
		Fermi energy at half filling.
		The arrows $q_a$ and $q_b$ indicate 
		interband scattering processes between 
		the ${k_y=\pi/2}$ and ${k_y=3\pi/2}$ bands 
		and the ${k_y=0}$ and ${k_y=\pi}$ bands,
		respectively.
	}
	\label{fig:4bands}
\end{figure}

Since our purpose is to study the influence of the interchain coupling
for a sufficient degree of frustration, 
we take the intrachain hopping as the unit of energy, i.e., set ${t=1}$,
and simultaneously vary the anisotropy, ${t_\perp}$, 
and the frustration, ${t'}$,
keeping the ratio of the two fixed to a fairly strong value:
${t'/t_{\perp}=-0.5}$.
We then tune ${t_\perp}$ between the limiting cases 
of uncoupled Hubbard chains at ${t_{\perp} = 0}$ 
and the isotropic case at ${t_{\perp} =1}$.
We work at half filling,
where we note that the Hamiltonian~\eqref{eqn:hubbard} 
is symmetric with respect to the sign of ${t'/t_{\perp}}$ 
if a particle-hole transformation is carried out,
so that only the absolute value of ${t'/t_{\perp}}$ is relevant.
We take ${t'/t_{\perp}}$ to be negative here 
in order to retain the form of the band structure in Fig.~\ref{fig:4bands} 
and to maintain notational consistency with other work.

In the present study, 
we concentrate on cylinders with a width of four lattice sites, ${L_y=4}$.
In the infinite-cylinder-length limit,
the four bands ${\varepsilon_j(k_x)}$ 
for transverse momentum ${k_y=j\,\pi/2}$ read
\begin{align}
	\varepsilon_0(k_x) & = 
	- 2\, (t - t_{\perp}) \, \cos k_x 
	- 2\, t_{\perp}  \,, \nonumber
	\\
	\varepsilon_{1}(k_x) = \varepsilon_{3}(k_x) & = 
	- 2\, t \, \cos k_x  \,,\nonumber
	\\
	\varepsilon_{2}(k_x) & = 
	- 2\, (t+t_{\perp}) \, \cos k_x 
	+ 2\, t_{\perp} \,.
	\label{eqn:4bands}	
\end{align}
In Fig.~\ref{fig:4bands}, the band structure is depicted for
${t_{\perp}=0.1}$, ${t_{\perp}=0.5}$, and ${t_{\perp}=0.9}$
with the Fermi level for the noninteracting case at half filling indicated.
We denote the Fermi points within each band by ${\bf k}^{\rm F}_j$.
For the case of weak interchain coupling, Fig.~\ref{fig:4bands}(a),
the four bands approach each other,
and the longitudinal Fermi momenta, $k^{\rm F}_{j\,x}$, 
converge towards ${\pm \pi /2}$ for ${t_\perp\rightarrow0}$.
In the ${t_{\perp}=0}$ limit, 
${k^{\rm F}_{j\,x}=\pm\pi/2}$ for ${j=0,\ldots,3}$,
the Fermi surface becomes nested, 
and intraband umklapp processes with momentum transfer of ${(2\,\pi,0)}$
are allowed within all four bands.
For intermediate interchain coupling, Fig.~\ref{fig:4bands}(b),
the Fermi points deviate from ${\pi/2}$,
and intraband umklapp processes are not permitted.
Still, interband umklapp processes 
utilizing all four bands with a total momentum transfer of
${\bf k}^{\rm F}_{0}+{\bf k}^{\rm F}_1+{\bf k}^{\rm F}_2+{\bf k}^{\rm
  F}_3={(2\,\pi,2\,\pi)}$ remain possible.
Note that this is possible because
at half filling ${\sum_{j=0}^3 k^{\rm F}_{j\,x}=2\,\pi}$, 
assuming that all bands are partially filled.
The scattering vectors involved,
${q_a:\,k^{\rm F}_1\rightarrow -k^{\rm F}_3}$ and 
${q_b:\,k^{\rm F}_0\rightarrow -k^{\rm F}_2}$, 
are depicted as gray dashed and solid arrows in Fig.~\ref{fig:4bands}(b).
Note that other interband umklapp processes involving all four Fermi bands
are also allowed; 
however, we only show the scattering vectors $q_a$ and $q_b$
since our DMRG results (see Sec.~\ref{sec:dmrg_results})
indicate that the relevant processes have transverse momentum $\pi$.
For strong interchain coupling, Fig.~\ref{fig:4bands}(c),
the ${k_y=0}$ band becomes flat and completely filled, 
only three bands remain active,
and no umklapp processes are allowed.

The transition between the latter scenarios
shown in Figs.~\ref{fig:4bands}(b) and~\ref{fig:4bands}(c),
i.e., the exact point where ${\varepsilon_0(k_x)}$ 
has a maximum at ${k^{\rm F}_{0\,x}=\pi}$, marks a Van Hove singularity.
In the two-dimensional case, this transition corresponds to the point where 
the noninteracting Fermi surface undergoes a topological change,
changing from an open to a closed surface,
as depicted in Fig.~\ref{fig:hubbard}(b).
Note that the critical interchain hopping strength
for the width-$4$ cylinder is ${t_{\perp}\approx0.707}$,
somewhat displaced from that of the two-dimensional case, 
where the singularity is at ${t_{\perp} \approx 0.622}$.

\section{DMRG methods}
\label{sec:DMRG}

We apply the DMRG within a hybrid-space representation
that is composed of a real-space representation in the longitudinal 
and a momentum-space representation in transverse direction,
as was recently introduced 
for the fermionic Hofstadter model~\cite{motruk2016hybridDMRG} and the
two-dimensional Hubbard model~\cite{ehlers2017hybridSpaceDMRG}.
In this representation, the transverse momentum ${k_y}$ is conserved; 
this can be utilized to speed the DMRG algorithm up significantly.
The retention of a real-space representation in the longer spatial
direction makes it possible to avoid
the undesirable volume-law scaling 
of the entropy~\cite{ehlers2015MomentumSpaceDMRG} that severely
restricts the applicability  
of the pure momentum-space DMRG~\cite{xiang1996momentumDMRG}.
Our implementation~\cite{ehlers2017hybridSpaceDMRG} is based on the
matrix product state (MPS) 
formulation of the DMRG~\cite{schollwock2011MPS}.
The utilization of conserved transverse momentum, total charge, 
and total spin quantum numbers enables us to push the dimension of the
virtual bonds of the MPS, $m$,  to up to ${35\,000}$ states; 
a detailed analysis of
the performance of the hybrid-space DMRG for the two-dimensional Hubbard model
is contained in Ref.~\cite{ehlers2017hybridSpaceDMRG}.

Despite the enhanced performance of the hybrid-space formulation,
the DMRG can still suffer from convergence difficulties.
First, the convergence of the DMRG, both in the real-space
and in the hybrid-space formulations, is exponentially costly in
system width due to the entropy area law~\cite{eisert2010areaLaw}. 
Therefore, we study the system at a relative small width
of ${L_y=4}$, i.e., four coupled chains.
For isolated parameter points, we will show results for ${L_y=5}$ in
order to roughly determine how the behavior changes 
as the number of chains is increased.
Calculations for ${L_y=5}$, for which convergence to the qualitatively
consistent ground state only occurs for ${m \gtrsim 20\,000}$, are
prohibitively expensive to carry out for the full parameter space
treated here.

Second, 
the algorithm sometimes gets trapped in a metastable state, 
which then, for calculations of finite accuracy, converges as the
apparent ground state of the system.
Such behavior is often found for parameter values close to a
discontinuous transition.
In order to minimize such problems and define the phase boundaries 
as accurately as possible,
we use a variation of the standard DMRG ground-state search:
we start DMRG calculations for two sets of model parameters
located on either side of, but sufficiently far from, 
a discontinuous phase transition.
We initialize both calculations by 
following the standard ground-state search procedure,
in which the model parameters are kept fixed
and $m$ is increased during the initial sweeps.
Once sufficient convergence has been attained for a particular
starting point, we fix $m$, continue sweeping,
and alter the model parameters in small steps towards the other phase.
As the system goes through a discontinuous
phase transition, we typically obtain a hysteresis effect and 
can then track the ground states of both phases into the parameter
regime of the other phase.
By determining the exact point at which the energies of the two states
cross, we obtain the phase boundary with maximal accuracy.

\begin{figure}
	\includegraphics[width=8.6cm]{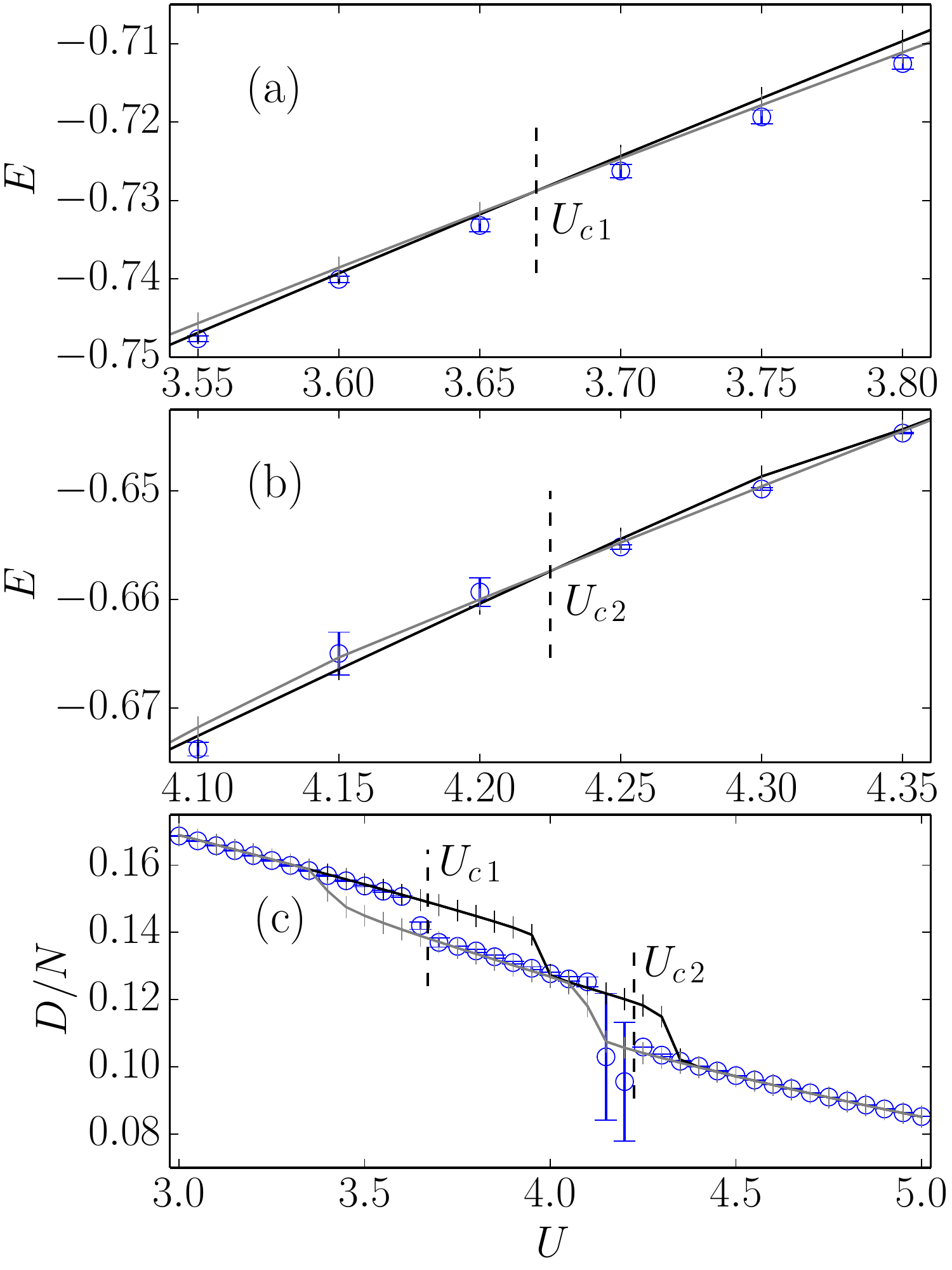}
	\caption{
		(Color online)
		Determination of the discontinuous phase transitions
        at ${t_{\perp}=0.7}$ and ${t'/t_{\perp}=-0.5}$ on
        a ${16{\times}4}$ lattice. 
		Panels (a) and (b) show the energy density ${E}$ as a
        function of ${U}$  with level crossings at
        ${U_{c\,1}}$ and ${U_{c\,2}}$;
		(c) shows the double occupancy per site, ${D/N}$, as a
        function of $U$, with $U_{c\,1}$ and $U_{c\,2}$,
        determined from (a) and (b), respectively, indicated.
		The black and gray solid lines show 
		right-moving and left-moving DMRG calculations,
		where $U$ is changed in steps of ${\Delta U=\pm 0.05}$ 
		at the beginning of every second sweep
		after the initial sweeps, respectively.
		The circles show results obtained by 
		the standard ground-state search procedure, 
		which are extrapolation to zero discarded weight.
		The error bars indicate the difference between
		extrapolated and non-extrapolated values.
	}
	\label{fig:doubleTracking}
\end{figure}
As an example, Fig.~\ref{fig:doubleTracking} shows calculations 
at fixed ${t_{\perp}=0.7}$ as a function of ${U}$. 
In this case, we find two discontinuous phase transitions 
at the critical interaction strengths ${U_{c\,1}}$ and ${U_{c\,2}}$.
The first two panels,
Figs.~\ref{fig:doubleTracking}(a) and~\ref{fig:doubleTracking}(b), 
show the energy-level crossings for the first and second transitions,
respectively.
These crossings correspond to finite jumps in the double occupancy 
${D=
\sum_{{\bf r}}\langle n_{{\bf r}\,\uparrow} n_{{\bf r}\,\downarrow} \rangle}$,
as can be seen in Fig.~\ref{fig:doubleTracking}(c).
This example shows that the hysteresis effect is present in the double
occupancy as well as in the energy.
Thus, we can determine the point of the transition 
and the height of the jump in the double occupancy accurately.
Results from standard DMRG calculations, carried out separately 
for every distinct value of $U$, are depicted as blue circles for comparison.

The procedure described above has both an advantage and disadvantages: 
the advantage is that convergence problems close to the phase boundary
are diminished because the critical initial phase of the DMRG
algorithm takes place  in a more stable region of the parameter space.
One disadvantage is that, since $m$ is fixed during the second phase 
of the calculation, the results cannot be extrapolated in the
truncation error without further effort.
Another is that the CPU time needed for a single calculation is comparably high 
because a large number of sweeps in which the maximum $m$ is kept must
be carried out.
In our calculations, we have used a fixed MPS bond dimension of ${m=10\,000}$ 
after the initial sweeps, which was large enough to converge the calculations 
and keep runtimes at an acceptable level.

In order to validate our results qualitatively, 
we have carried out additional calculations in which we used the
standard DMRG ground-state search at all points of parameter space.
For these calculations, we have used a maximum bond dimension 
between ${m=20\,000}$ and ${m=35\,000}$,
have varied the chain length,
and have extrapolated the results in the truncation error.

\section{variational cluster approximation}
\label{sec:VCA}

The variational cluster approximation (VCA)~\cite{dahnken2003VCA} is a quantum cluster technique~\cite{maier2005},
which is used to study strongly correlated electron systems with local
interactions and to investigate phases with broken
symmetries~\cite{dahnken2004}.
Its validity has been successfully demonstrated for the 
one-dimensional Hubbard model~\cite{dahnken2003VCA,balzer2008},
and it has been applied to two-dimensional
Hubbard systems~\cite{dahnken2004,aichhorn2006},  
allowing for both, the investigation of magnetically ordered 
phases~\cite{dahnken2004,tremblay2008frustratedHubbard,
yamada2013frustratedHubbard} 
and of the Mott
transition~\cite{balzer2009,lenz2016dimensionalCrossoverFrustratedHubbard}.

The VCA is based on Potthoff's self-energy functional
theory~\cite{potthoff2003, potthoff2003a, potthoff12}, 
in which the grand-canonical potential of the infinite system is written 
as a functional of the self-energy $\Sigma$,
\begin{equation}
	\Omega [\Sigma] = F [\Sigma] - \mathrm{tr} \log ( - G_0^{-1} + \Sigma ),
\end{equation}
where $F[\Sigma]$ denotes the Legendre-transformed Luttinger-Ward
functional~\cite{luttinger1960,potthoff2006a}, and 
${G_0=(\omega+\mu- \mathbf{t})^{-1}}$ is the noninteracting
single-particle Green's function.
At the physical self-energy $\Sigma_{\text{phys}}$, 
this functional is stationary as a function of the self-energy, 
i.e., ${\delta \Omega[\Sigma] / \delta
\Sigma\vert_{\Sigma_{\text{phys}}} = 0}$.
Since ${F[\Sigma]}$ is usually not known, 
one has to resort to approximations to determine the self-energy 
so that this stationary condition is fulfilled. 

In the VCA, a reference system consisting of decoupled clusters 
is used to calculate the self-energy functional.
This reference system, described by a Hamiltonian ${H^{\prime}}$, 
has the same interaction terms as the original system 
and the cluster self-energy can be computed exactly, 
for example, using exact diagonalization.
Since ${F[\Sigma]}$ is universal in the sense that it only depends 
on the interaction part of the Hamiltonian, both the original 
and the reference systems have the same Luttinger-Ward functional.
Therefore, the self-energy functional 
can be expressed in terms of the reference system as
\begin{equation}
	\Omega [\Sigma] = \Omega^{\prime} [\Sigma] + \mathrm{tr} \log ( - {G_0^{\prime}}^{-1} + \Sigma ) - \mathrm{tr} \log ( - G_0^{-1} + \Sigma ),
\end{equation}
where all cluster quantities are denoted by a prime.
Through variation of the one-body parameters ${\mathbf{t}^{\prime}}$ 
of the reference system, the stationary point of the functional 
can be determined because
\begin{equation}
 \frac{\delta\Omega[\Sigma]}{\delta \mathbf{t}^{\prime}} =
 \frac{\delta\Omega[\Sigma]}{\delta\Sigma} \,
 \frac{\delta\Sigma}{\delta\mathbf{t^{\prime}}}=0\,.
 \label{eqn:VCAvariation}
\end{equation}
The approximation of the VCA thus lies in restricting the variational space 
of all possible self-energies in Eq.~\eqref{eqn:VCAvariation}
to a limited set of cluster-representable self-energies $\Sigma^{\prime}$.

In order to account for phases with broken symmetry such as an
antiferromagnetic phase,
additional Weiss fields are added to the cluster Hamiltonian,
and their field strength is used as one of the variational parameters
to find the stationary point of ${\Omega[\Sigma]}$.
In this way, short-range correlations within the cluster are treated exactly,
but longer-ranged correlations are treated on a mean-field 
level~\cite{aichhorn2006}.
In order to properly describe the discontinuous (paramagnetic) Mott transition,
it was shown that it is necessary 
to include noninteracting bath sites in the reference cluster~\cite{balzer2009}.
This is permitted within self-energy functional theory because
additional couplings to noninteracting sites do not change the 
Luttinger-Ward functional, 
and the clusters thus are a valid reference system. 

\section{Results}
\label{sec:results}

In the following, we present our results for the DMRG,
Sec.~\ref{sec:dmrg_results}, and the VCA, Sec.~\ref{sec:vca_results}.
The DMRG results cover a detailed phase diagram
in the $t_{\perp}$ and $U$ plane 
and determine the static magnetic and electric properties of the phases.
For the VCA, we focus on four different values of $t_{\perp}$, 
for which we determine the position and nature 
of the metal-insulator transitions and compare them to the DMRG results.

\subsection{DMRG results}
\label{sec:dmrg_results}

\begin{figure}
	\includegraphics[width=8.6cm]{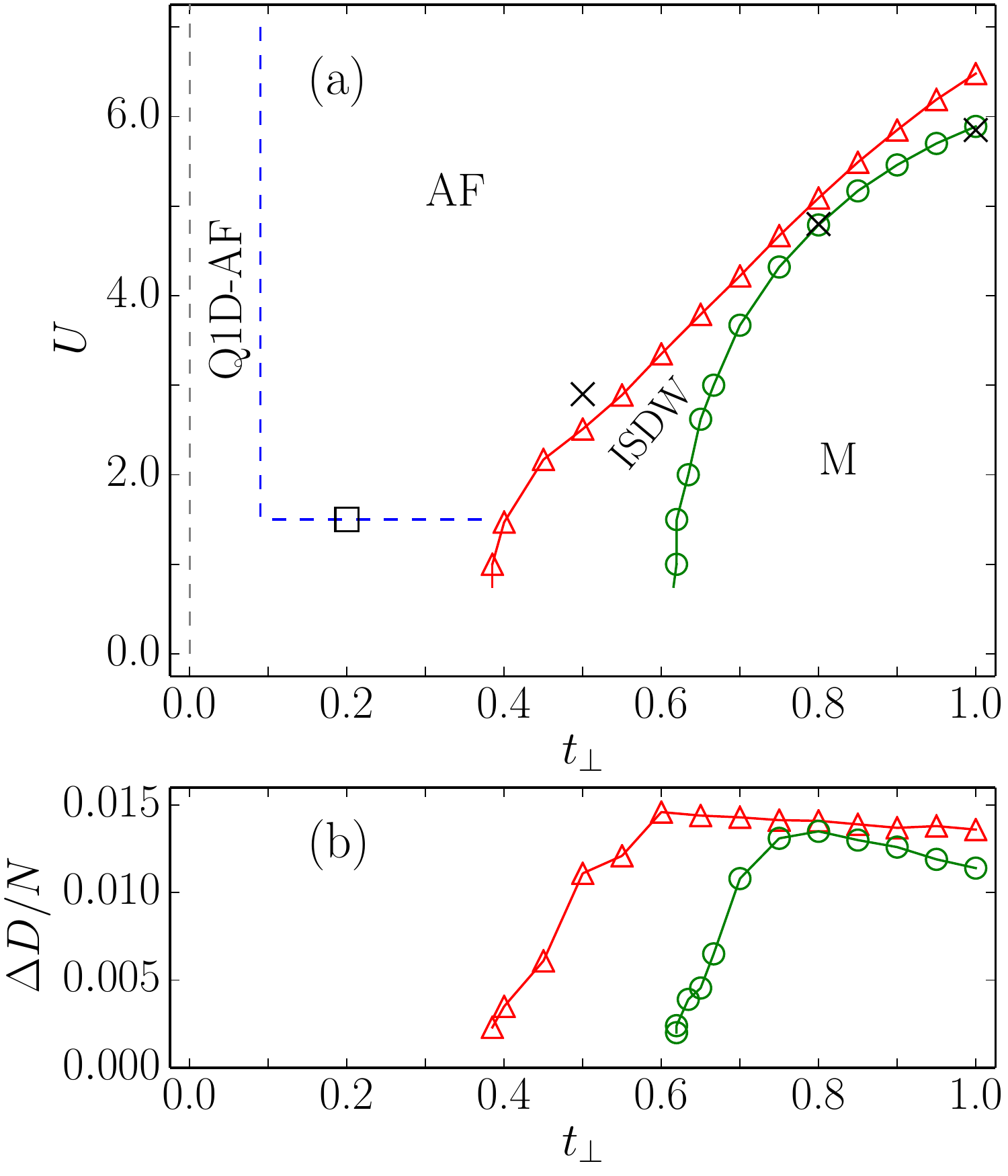}
	\caption{
	    (Color online) 
	    (a) DMRG phase diagram of the anisotropic half filled Hubbard model
	    with frustration ${t'/t_{\perp}=-0.5}$, 
	    obtained for width-$4$ cylinders,	    
	    including the VCA results for selected values of $t_{\perp}$.
	    The designations for the phases are:
	    metallic conductor (M), antiferromagnet (AF),
	    quasi-one-dimensional antiferromagnet (Q1D-AF),
	    and incommensurate spin-density wave (ISDW).
		Discontinuous phase transitions are depicted as solid lines.
	    The case of uncoupled chains, ${t_{\perp}=0}$, 
	    is marked by a gray dashed line, 
	    and corresponds to the purely one-dimensional antiferromagnet.
	    The transition points
	    between the metallic phase and the antiferromagnetic insulator
	    as seen by the VCA
	    (see Sec.~\ref{sec:vca_results}) are indicated:
	    black crosses mark discontinuous transitions
        for ${t_{\perp}=0.5}$, ${t_{\perp}=0.8}$, and ${t_{\perp}=1.0}$,
	    and the black square marks a continuous 
	    transition found for ${t_{\perp}=0.2}$.
	    (b) Jump in the double occupancy per site, ${\Delta D/N}$,
        along the transition lines, parametrized with $t_\perp$, 
        for the transition between the ISDW and the metallic phases 
        (green line with circles) 
	    and between the ISDW and the AF / Q1D-AF phases 
	    (red line with triangles),
	    obtained for the ${16{\times}4}$ lattice. 
	}
	\label{fig:phaseDiagram}
\end{figure}
We now use  the tracking method described in Sec.~\ref{sec:DMRG} 
to locate phase-transition lines in the $t_\perp$--$U$ plane 
for cylinders with ${16{\times}4}$ sites.
We find two phase-transition lines in the regime between ${t_\perp=0}$
and ${t_\perp=1}$, both of which are due to level crossing scenarios 
in the ground state for the entire range of $t_\perp$. 
The resulting phase diagram is depicted in Fig.~\ref{fig:phaseDiagram}.
Note that the transition lines are determined 
only for the ${16{\times}4}$ lattice.
In order to unambiguously determine the order of the phase transitions, 
a finite-size scaling analysis would be needed.
Due to the high computational costs, we refrain from doing this here.
However, as discussed further below in this section, 
we perform a finite-size scaling analysis for the electrical susceptibility,
and investigate in detail the behavior of the spin structure factor.
The behaviors of both quantities support that the lines of level crossings 
depicted in Fig.~\ref{fig:phaseDiagram} will indeed remain discontinuous
phase transitions in the infinite-length limit.

The phase diagram in Fig.~\ref{fig:phaseDiagram}(a) features a metallic phase, 
an antiferromagnetic phase (AF), 
a quasi-one-dimensional antiferromagnetic phase (Q1D-AF),
and an incommensurate spin-density-wave phase (ISDW) 
that separates the metallic and the antiferromagnetic phases.
Both transition lines are characterized by
finite jumps in the double occupancy, 
${\Delta D}$, Fig.~\ref{fig:phaseDiagram}(b), 
along the lines indicating the presence of discontinuous phase transitions.
The metallic phase extends from ${U=0.0}$ to ${U\approx 6.0}$ 
in the isotropic case ${({t_{\perp}=1})}$,
becomes narrower with decreasing ${t_{\perp}}$,
and vanishes at ${t_{\perp} \approx 0.6}$.
Note that close to the latter point, at ${t_{\perp} \approx 0.707}$,
a Van Hove singularity marks the transition between three and four active bands
in the band structure for width-$4$ cylinders; see Sec.~\ref{sec:model}.
The ISDW phase bounds the metallic phase and is present
for transverse hopping strength ${0.4 \lesssim t_{\perp} \leq 1.0}$.
These DMRG results for the phase diagram are compared to transitions 
found by the VCA at ${t_\perp = 0.5}$, $0.8$, and $1.0$. 
As can be seen in Fig.~\ref{fig:phaseDiagram}, 
discontinuous transitions from a metallic to an AF phase 
are found for large values of $t_\perp$, 
which agree very well with the locations of the transitions 
identified by the DMRG. 
At smaller values of $t_\perp$, 
a continuous transition is seen, 
which is difficult to capture using the DMRG. 
These results and the comparison to the DMRG phase diagram 
will be discussed in detail in the following and in Sec.~\ref{sec:vca_results}.

\begin{figure}
	\includegraphics[width=8.6cm]{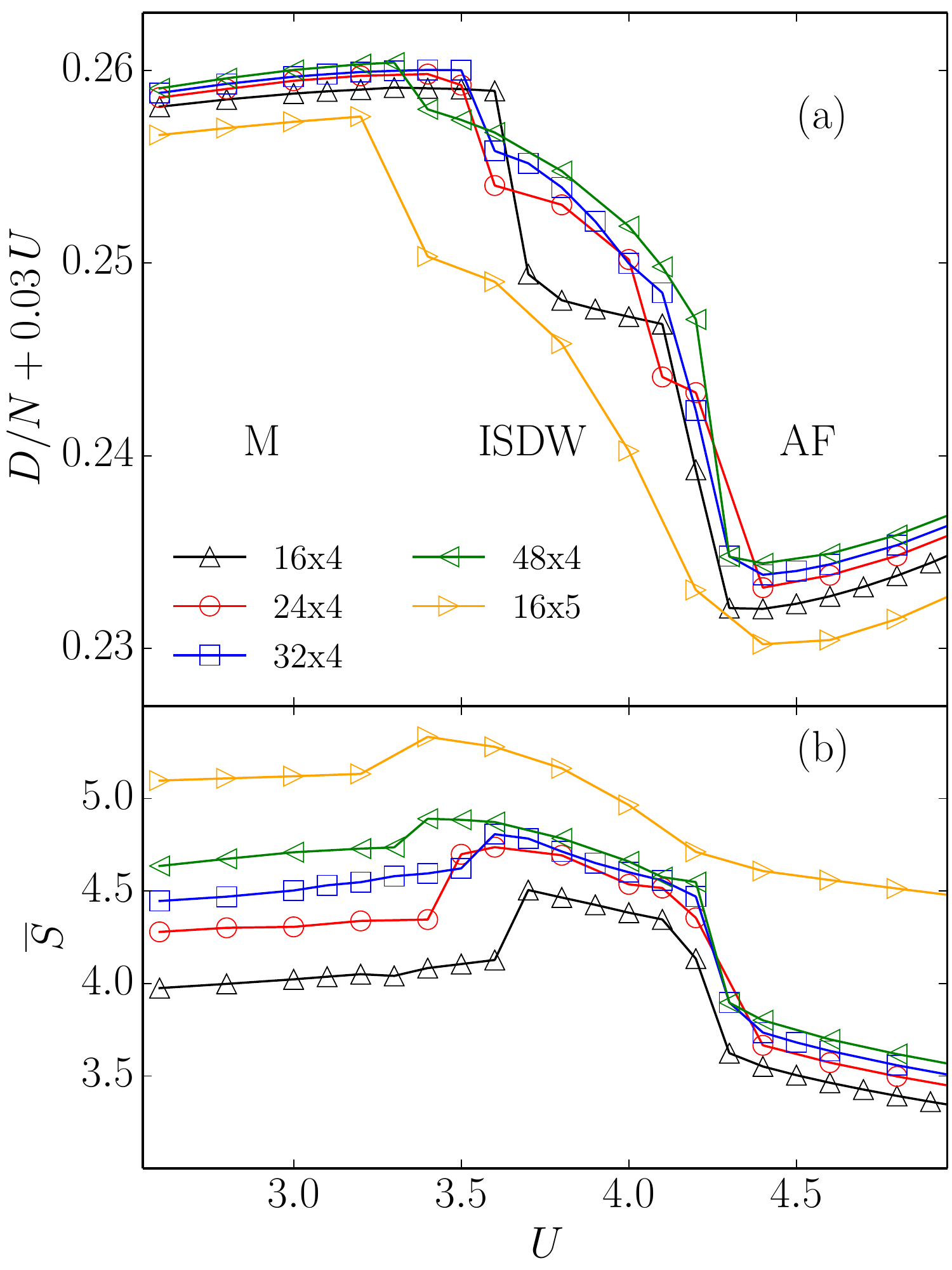}
	\caption{
		(Color online)     
		(a) Double occupancy per site, $D/N$, 
		and (b) averaged von Neumann entropy, 
		$\overline{S}$ [Eq.~\eqref{eq:Sbar}], 
		for different system sizes ${L_x{\times}L_y}$
		at ${t_{\perp}=0.7}$.
		Note that $D/N$ is shifted by ${0.03U}$ 
		to improve the visibility of the jumps
		between the indicated phases.
	}	
	\label{fig:doubleEntropy}
\end{figure}
Figure.~\ref{fig:doubleEntropy}
shows the behavior of the double occupancy and the block entropy
at the two transitions between the metallic and the ISDW phases
and between the ISDW and AF phases at fixed ${t_\perp=0.7}$
for width-4 cylinders with different lengths 
and for the width-5 cylinder of length 16.
In Fig.~\ref{fig:doubleEntropy}(a), 
the double occupancy is shifted by ${0.03U}$ 
to compensate for the linear growth in $U$ 
and to emphasize the jumps between the phases.
For width-4 cylinders, the jump between the ISDW and AF phases
seems to be stable as a function of the system length.
At the transition between the ISDW and the metallic phase,
the jump is present for all system lengths,
but shrinks somewhat with increasing $L_x$.
Thus, from the behavior of
the double occupancy alone, 
we cannot rule out that this transition becomes continuous
in the infinite-length limit.
In order to confirm this,
one would need to perform a scaling analysis, 
which cannot be done reliably with the data set available.

The discontinuous behavior is also marked by a discontinuity in the entropy.
Figure~\ref{fig:doubleEntropy}(b) shows the averaged von Neumann entropy,
\begin{equation}
	\overline{S} = \frac{1}{N}\sum_i S(i) \, ,
	\label{eq:Sbar}
\end{equation}
where ${S(i)=-\mathrm{tr}(\rho_i \ln \rho_i)}$ 
is the von Neumann, entanglement entropy~\cite{amico2008entanglement}
calculated for a bipartite partitioning of the system at
the $i$th MPS bond and $\rho_i$ the corresponding reduced density
matrix of one of the subsystems.
(Note that the state obtained within the DMRG is a pure state, usually an
approximation to the ground state of the system.)
Again, for width-4 cylinders, the jump is stable 
for the transition between the ISDW and AF phases,
but the data does not allow for a definite statement
on the nature of the transition between the metallic and ISDW phases.

For the ${16{\times}5}$ lattice, the signatures of both transitions 
are evident in the double occupancy and in the entropy.
For width 5, a scaling in system length is not possible 
due to the high computational costs.

\begin{figure}
	\includegraphics[width=8.6cm]{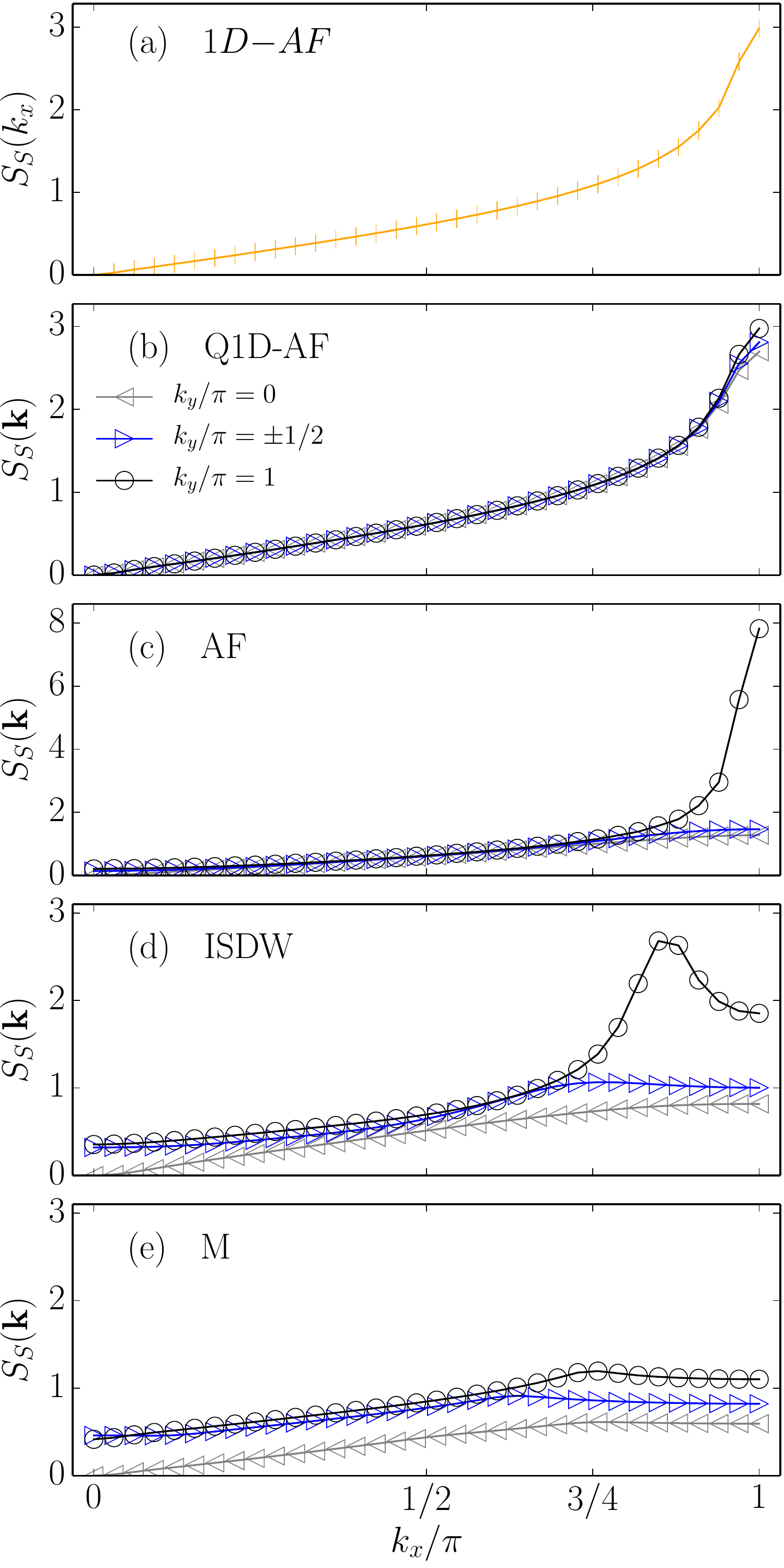}
	\caption{
		(Color online)
		(a) Static spin structure factor ${S_{S}(k_x)}$
		[Eq.~\eqref{eqn:spinStructFactor}]
		of the one-dimensional 32-site Hubbard model at ${U=4.0}$.
		(b)--(e) Static spin structure factor ${S_{S}({\bf k})}$
		of ${32{\times}4}$ Hubbard cylinders for ${U=4.0}$ 
		with varying interchain hopping:
 		(b) ${t_{\perp}=0.075}$,
		(c) ${t_{\perp}=0.5}$,
 		(d) ${t_{\perp}=0.7}$,
 		and (e) ${t_{\perp}=0.9}$.
 	}
	\label{fig:spinStructure}
\end{figure}
We classify the magnetic ordering
of the AF, Q1D-AF, and ISDW phases by 
examining the static structure factor
\begin{equation}	
	S_{S}({\bf k}) = 
	\frac{1}{N} \sum_{\bf{r} \, {\bf r}'} 
	e^{\rm{i} {\bf k} ({\bf r} - {\bf r}')} \, S({\bf r},{\bf r}')
	\label{eqn:spinStructFactor}
\end{equation}
of the static spin correlations 
\begin{equation}
	S({\bf r},{\bf r}') =  \langle S_z({\bf r}) \,  S_z({\bf r}') \rangle 
	- \langle S_z({\bf r}) \rangle\langle S_z({\bf r'}) \rangle
\end{equation}
with 
	$ {S_z(x,y) = } \,  
	{c^{\dagger}_{x\,y\,\uparrow} \, 
	c^{\vphantom{\dagger}}_{x\,y\,\uparrow}}
	-{c^{\dagger}_{x\,y\,\downarrow} \, 	
	c^{\vphantom{\dagger}}_{x\,y\,\downarrow} }$.
Since we have
open boundary conditions in the  $x$ direction,
we use particle-in-a-box eigenmodes for the transformation in that
direction to approximate momentum modes,
as in Ref.~\cite{hager2005stripe}.
Figure~\ref{fig:spinStructure} depicts the structure factor
with panel (a) showing the result for a single chain and
panels (b)--(e) following a cut through the
phase diagram at constant ${U=4.0}$ with increasing interchain coupling.
For weak ${t_\perp=0.075}$, Fig.~\ref{fig:spinStructure}(b),
${S_{S}({\bf k})}$ has one peak at ${k_x=\pi}$ 
in all four transverse momentum branches,
but the form of ${S_{S}({\bf k})}$ only varies weakly with
transverse momentum, $k_y$.
The shape of ${S_{S}({\bf k})}$ in the longitudinal direction 
has almost exactly the same form as that of the one-dimensional case, 
Fig.~\ref{fig:spinStructure}(a), especially for the ${k_y=\pi}$ branch.
For larger ${t_\perp}$, the two-dimensional antiferromagnetic phase 
is characterized by a very strong peak in
${S_{S}({\bf k})}$ at ${{\bf k}=(\pi,\pi)}$, 
which contrasts with much weaker peaks at ${k_x=\pi}$ 
in the other $k_y$ channels, 
as can be seen in Fig.~\ref{fig:spinStructure}(c) 
for a representative point at ${t_\perp=0.5}$. 

\begin{figure}
	\includegraphics[width=8.6cm]{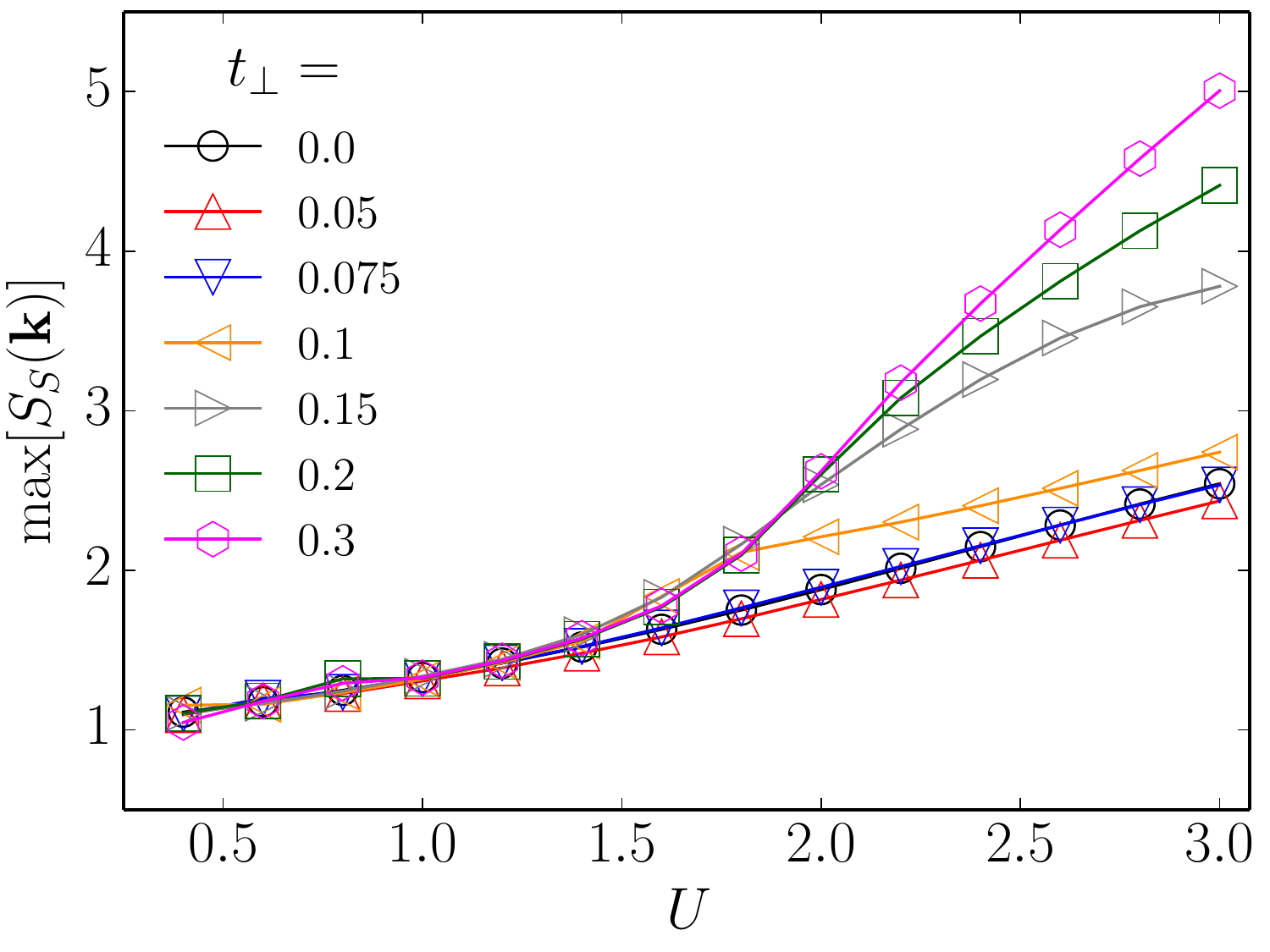}
	\caption{
		(Color online) 
		Peak height of the spin structure factor, ${\max[S_S({\bf k})]}$ 
		[see Eq.~\eqref{eqn:spinStructFactor}],
		as a function of the interaction strength ${U}$
		for the one-dimensional Hubbard chain ${(t_{\perp}=0.0)}$
		and for ${32{\times}4}$ Hubbard cylinders.
	}
	\label{fig:spinPeak}
\end{figure}
The $U$-dependence of the height of the peak at ${{\bf k}=(\pi,\pi)}$ 
is displayed in Fig.~\ref{fig:spinPeak}.
For small to moderate ${t_\perp\lesssim 0.3}$,
the height of the peak at ${{\bf k}=(\pi,\pi)}$ grows as $U$ is increased.
For all ${t_\perp\lesssim 0.3}$, the growth follows that of
the one-dimensional case, ${t_\perp=0.0}$, up to ${U\approx 1.5}$.
Above ${U \approx 1.5}$, 
the peak heights continue to follow the one-dimensional case for
smaller $t_\perp$ (${t_\perp\lesssim 0.1}$), but for
${0.1 \lesssim t_\perp \lesssim 0.3}$, the peak height grows
significantly more strongly with $U$ than the one-dimensional case.
This is an indication of the crossover to a two-dimensional
antiferromagnetic phase.
The crossover line determined in this way is marked by the blue dashed line in 
Fig.~\ref{fig:phaseDiagram}(a).

\begin{figure}
	\includegraphics[width=8.6cm]{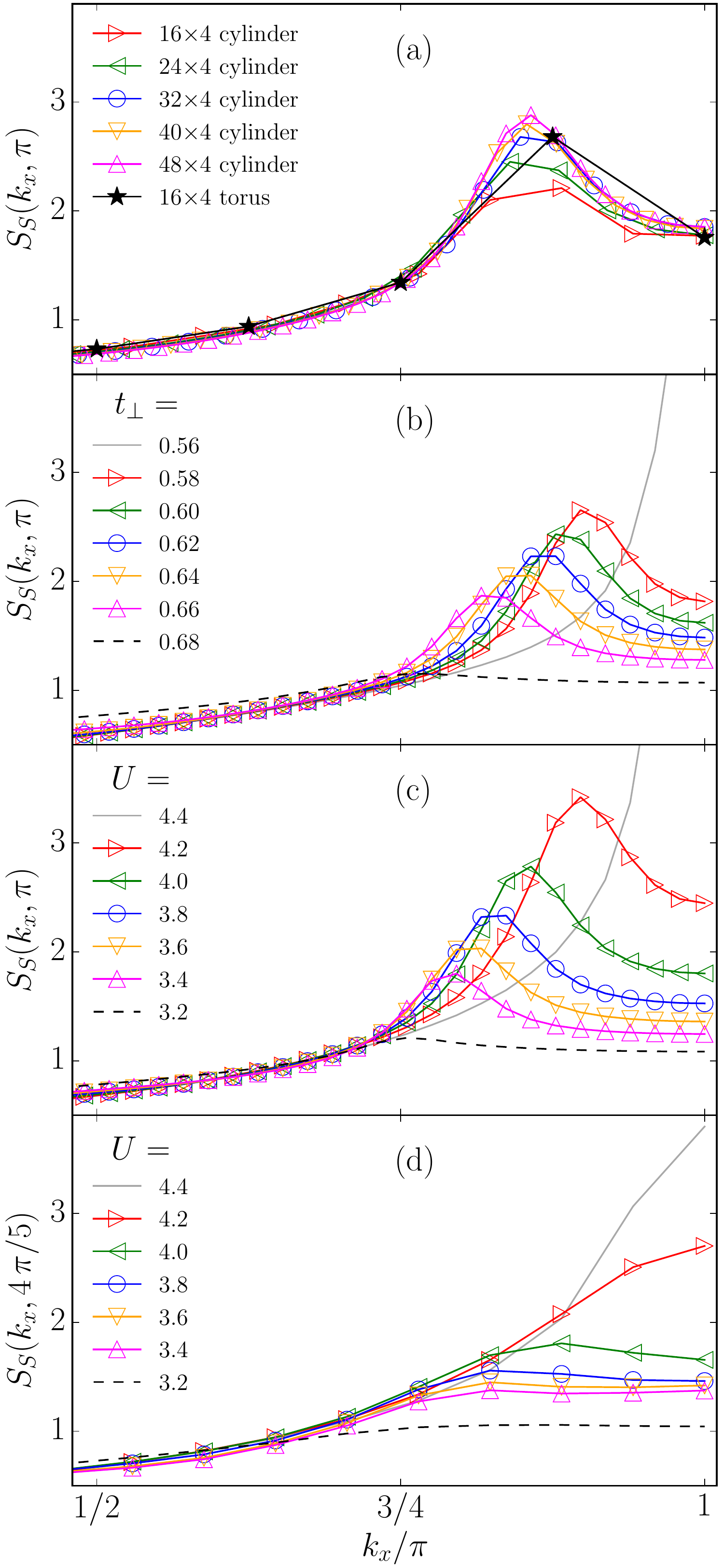}
	\caption{
		(Color online) 
		Spin structure factor ${S_S(k_x,\pi)}$
		[Eq.~\eqref{eqn:spinStructFactor}] in the ISDW phase
		for the indicated model parameters and 
		(a) for fixed ${t_{\perp}=0.7}$ and ${U=4.0}$
		for width-4 cylinders as a function of system length ${L_x}$ 
		and for a $16{\times}4$ lattice with toroidal lattice topology,
		(b) for ${48{\times}4}$ cylinders
		as a function of  ${t_{\perp}}$ at fixed ${U=3.0}$, 
		and (c) for ${48{\times}4}$ cylinders
		as a function of ${U}$ at fixed ${t_{\perp}=0.7}$.
		(d) ${S_S(k_x,4\,\pi/5)}$ for ${16{\times}5}$ cylinders 
		as a function of ${U}$ at fixed ${t_{\perp}=0.7}$.		
		In panels (b)--(d),
		the black dashed and gray solid lines depict
		${S_S(k_x,\pi)}$ within the metallic and
        antiferromagnet phases, respectively.
	}
	\label{fig:sdw}
\end{figure}
The ISDW phase, Fig.~\ref{fig:spinStructure}(d),
is characterized by an incommensurate ordering wave vector 
	${{\bf Q}^*=(\pm q^*,\pi)}$,
where the longitudinal wave vector of the peak, $q^*$, 
depends on ${U}$ and ${t_{\perp}}$. 
This incommensurate structure
corresponds to antiferromagnetic spin correlations in real space with 
a sine-shaped modulation in the longitudinal direction,
where the envelope has a wavelength of ${2\,\pi/(\pi-q^*)}$.
Figure~\ref{fig:sdw} shows plots of ${S_S(k_x,\pi)}$ within the ISDW phase
for varying $L_x$, ${t_{\perp}}$, and ${U}$.
In the metallic phase, depicted for ${t_\perp=0.9}$ and ${U=4.0}$ in
Fig.~\ref{fig:spinStructure}(e),
the structure factor ${S_S({\bf k})}$ lacks a distinct peak in all 
transverse momentum sectors.

Figure~\ref{fig:sdw}(a) shows that the wave vector of the
incommensurate peak, $q^*$, is stable as a function of system length $L_x$
for $L_x$ larger than the corresponding wavelength.
All but one of the curves shown are calculated with open
boundary conditions in the $x$ direction.
The additional curve is calculated on a $16\times4$ lattice 
with periodic boundary conditions 
(the largest periodic lattice size for which we obtained good convergence).
While the finite-size effects and the placement of momentum points are
different for periodic than for open boundary conditions 
(in particular, the momentum points are more widely spaced), 
the values of ${S_S(k_x,\pi)}$ and the approximate position 
of the peak are consistent with the open-boundary-condition results, 
showing that the incommensurate peak structure 
is not an artifact of the boundary conditions.

As ${t_{\perp}}$ and ${U}$ are varied, 
Figs.~\ref{fig:sdw}(b) and~\ref{fig:sdw}(c),
$q^*$ changes, ranging between ${3 \, \pi/4}$ and $\pi$, with $q^*$
moving towards $\pi$, i.e., with the wavelength of the modulation becoming
longer, as the AF phase is approached, and  $q^*$
moving towards ${3 \, \pi/4}$ (shorter modulation wavelength) as the
metallic phase is approached.
Note that this movement of the peak position,  $q^*$,
can only be observed for systems of sufficient length;
if $L_x$ is too small,
the wavelength can get locked in at a particular discrete
value due to the boundary conditions.
This effect causes the large discrepancy
in the double occupancy in Fig.~\ref{fig:doubleEntropy}(a)
between system lengths $16$, $24$, and $32$.

In order to roughly gauge the effect of cylinder width, we 
display the spin structure factor for the ${16{\times}5}$ cylinder in
Fig.~\ref{fig:sdw}(d).
Note that the antiferromagnetic ordering is frustrated by the odd system width,
and, for the periodic transverse boundary conditions used here,
transverse momentum ${k_y=\pi}$ is not available, 
so that we take the closest point, ${k_y=4\,\pi/5}$.
As can be seen, the incommensurate peak characterizing the ISDW phase 
is still present, but less distinct for ${U=3.4\text{--}4.2}$.
At ${U=4.4}$, the peak is at ${k_x=\pi}$, 
as expected for the antiferromagnetic phase, and, at ${U=3.2}$, 
the form is as expected for the metallic phase.
Note that the peaks are also weakened due to
the small system length; compare to Fig.~\ref{fig:sdw}(a).

\begin{figure}
	\includegraphics[width=8.6cm]{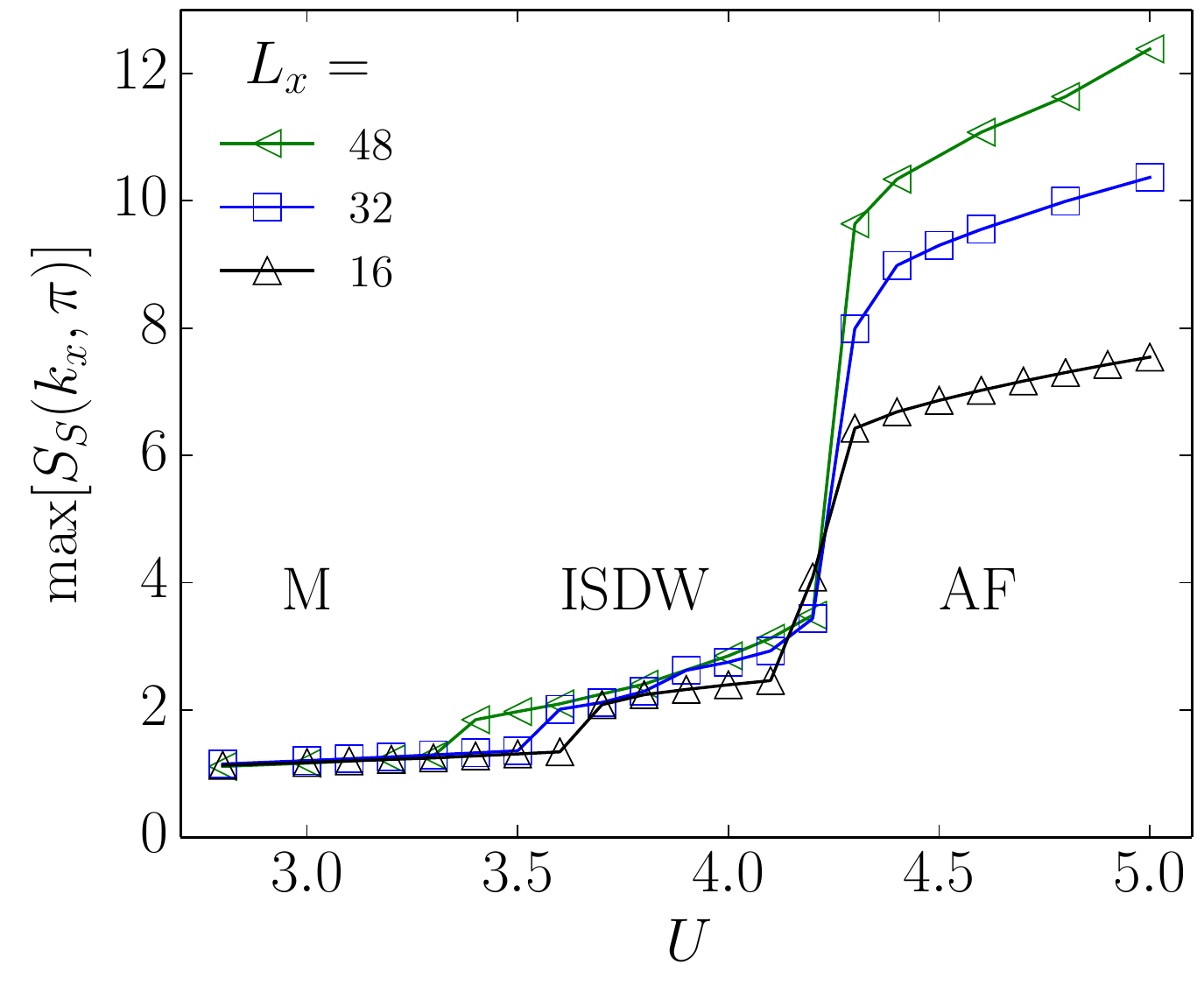}
	\caption{
		(Color online) Peak height of the spin structure factor [Eq.~\eqref{eqn:spinStructFactor}],
		${\max[S_S(k_x,\pi)]}$,
		for width-4 cylinders of the indicated lengths
		at ${t_{\perp}=0.7}$ as a function of $U$,
		with the metallic, ISDW, and AF phases indicated.
	}
	\label{fig:ssf}
\end{figure}

The static spin structure factor can also be used
to characterize the transitions between the ISDW and the metallic and
between the metallic and the AF phases.
The peak height and position changes smoothly within the ISDW phase, 
but changes abruptly at the transition points toward the  metallic 
and antiferromagnetic phases;
compare Fig.~\ref{fig:spinStructure}(d) with
Figs.~\ref{fig:spinStructure}(e) and 
\ref{fig:spinStructure}(c), respectively.
These discontinuous transitions are most evident in the behavior of
the peak height of the structure factor,
as plotted for width-4 cylinders of different lengths in Fig.~\ref{fig:ssf}.
It can be seen that, although the position of the transition
between the metallic and the ISDW phases changes,
the size of the jump in the peak height
at the transition remains stable as a function of $L_x$.
For the transition between the AF and the ISDW phase,
the position of the peak is stable and the height of the jump increases with 
system length.
In the AF phase, the peak height increases with cylinder length.
However, we note that the scaling is quasi-one-dimensional because we scale
with system length only.
In one dimension, long-range antiferromagnetic order cannot occur; the
correlations can decay at most with a power law \cite{giamarchi2003}.
Increase of the peak height with system size will only occur if the
power-law correlations fall off more slowly than $1/x$ (with
logarithmic scaling occurring at $1/x$).
This picture is consistent with the scaling of the AF peak in
Fig.~\ref{fig:ssf}, which seems to grow sublinearly with cylinder
length.
In the ISDW phase, there is no significant scaling of the peak height
with cylinder length, especially for the two larger $L_x$ values.
Thus, if power-law decay is present, it must be more rapid than $1/x$,
and we cannot differentiate between such a decay and exponential decay
with a relatively long decay length, i.e., between a weakly critical
and a weakly gapped phase.

\begin{figure}
	\includegraphics[width=8.6cm]{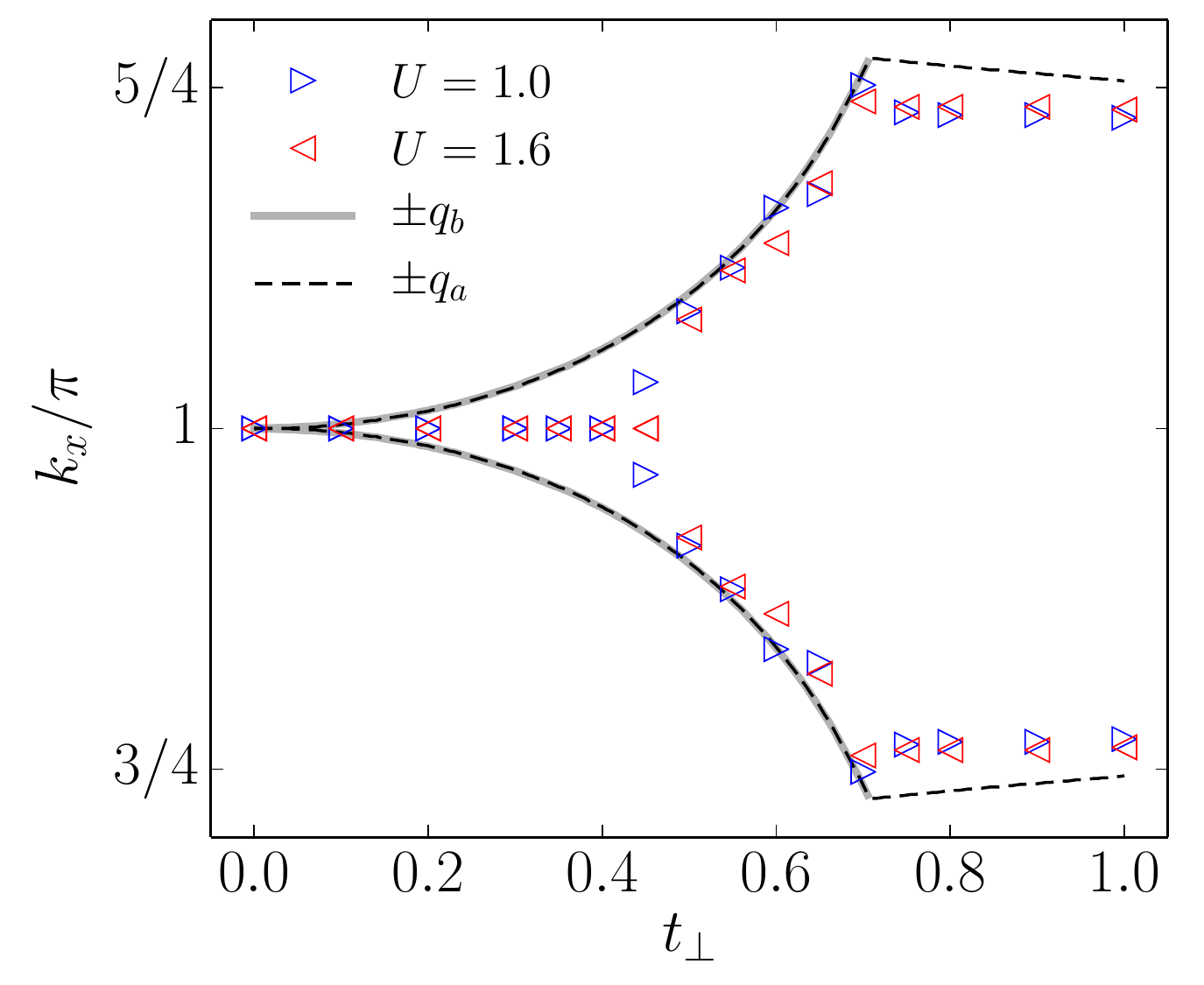}
	\caption{
		(Color online) 
		Symbols:
		Position of the maximum of ${S_S(k_x,\pi)}$ [Eq.~\eqref{eqn:spinStructFactor}]
		for ${48{\times}4}$ cylinders at ${U=1.0}$ and  ${U=1.6}$
		as a function of $t_\perp$.
		Lines: 
		Scattering wave vectors ${q_a=k^{\rm F}_{0\,x}+k^{\rm F}_{2\,x}}$
		and ${q_b=k^{\rm F}_{1\,x}+k^{\rm F}_{3\,x}}$; see Fig.~\ref{fig:4bands}.
	}
	\label{fig:ssf2}
\end{figure}

For sufficiently weak interaction strength,
one might expect that longitudinal wave  vector $q^*$ 
of the peak of the spin structure factor
would be determined by wave vectors characteristic of relevant
scattering processes in a weak-coupling picture.
Such scattering wave vectors are determined by
the ${U=0}$ band structure, Eq.~\eqref{eqn:4bands}, 
as described in Sec.~\ref{sec:model}, 
and are depicted in Fig.~\ref{fig:4bands}.
Figure~\ref{fig:ssf2} shows the position of the maximum of ${S_S(k_x,\pi)}$
as a function of $t_\perp$ for ${U=1.0}$ and ${U=1.6}$.
For small $t_\perp$, the maximum is at transverse momentum $\pi$.
Within the ISDW phase, 
the single maximum splits into two maxima symmetric to $\pi$,
which move towards ${(1\pm1/4)\pi}$ as $t_\perp$ is increased,
and finally, in the metallic phase,
the positions of the maxima remain stable at around ${(1\pm1/4)\pi}$.
For comparison,
scattering wave vectors ${q_a=k^{\rm F}_{0\,x}+k^{\rm F}_{2\,x}}$ and
${q_b=k^{\rm F}_{1\,x}+k^{\rm F}_{3\,x}}$,
defined as in Fig.~\ref{fig:4bands}(b), are depicted.
As can be seen, the DMRG results for both ${U=1}$ and ${U=1.6}$
are in good agreement with the scattering wave vectors
given by the ${U=0}$ band structure,
including the signature of the Van Hove singularity at ${t_\perp\approx0.7}$.

\begin{figure}
	\includegraphics[width=8.6cm]{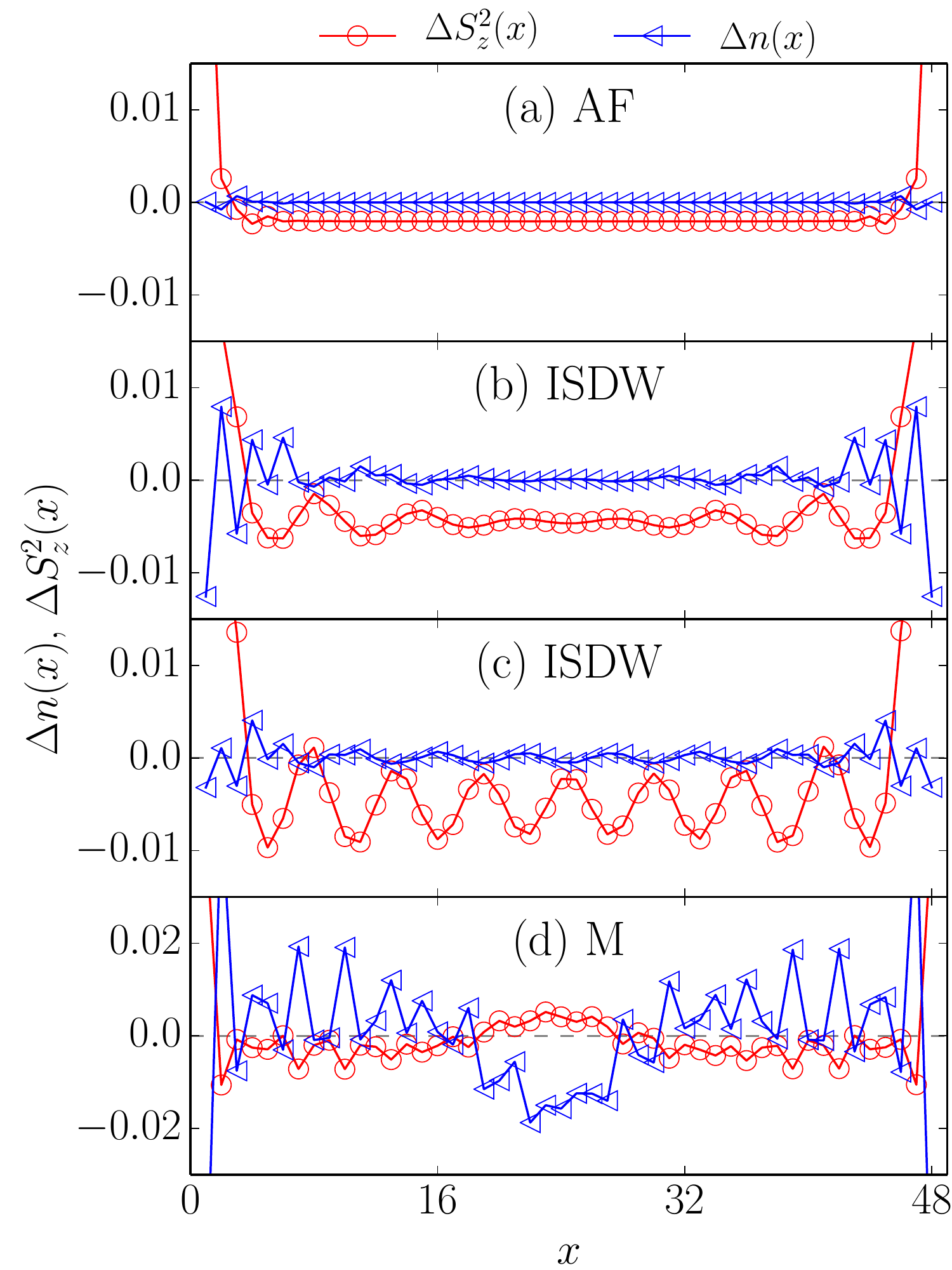}
	\caption{
		(Color online) 
		Fluctuations of the spin and charge density,
		${\Delta S_z^2(x)}$ and ${\Delta n(x)}$ 
		for system size ${48{\times}4}$
		for selected points in the phase diagram:
		(a) ${t_{\perp}=0.7}$ and ${U=5.0}$,
		(b) ${t_{\perp}=0.7}$ and ${U=4.0}$,
		(c) ${t_{\perp}=1.0}$ and ${U=6.0}$,
		and (d) ${t_{\perp}=0.7}$ and ${U=3.0}$.
	}
	\label{fig:scdw}
\end{figure}
Since translational invariance is broken 
in the longitudinal direction by the boundary conditions, the
structure of the phases can be 
seen in the behavior of local quantities, in particular, the
deviation of the local charge density, 
	${\Delta \, n (x)} = 
	{L_y^{-1}\sum_{y} \langle n(x,y) \rangle - 1}$, 
from its average value ${n=1.0}$ 
and the deviation of the local $z$-spin moment squared,
	${\Delta \, S_z^2(x)} = 
	{L_y^{-1}\sum_{y} \langle S_z^2(x,y) \rangle} 
	- {S_z^2(x,y)}$, 
from its average value 
	${S_z^2= N^{-1}\sum_{\bf r} \langle S_z^2(\bf r) \rangle}$
as a function of $x$, the position in the longitudinal direction. 
In the AF phase, as depicted in Fig.~\ref{fig:scdw}(a), spatial
fluctuations in both the charge and spin densities are strongly
suppressed, with the end effects from the open boundaries
damping out very quickly.
Thus, the dominant antiferromagnetic correlations strongly suppress
all spatial fluctuations in the charge and spin density.
Note that the average value ${S_z^2 = 0.825}$ 
of the $z$-spin moment squared is relatively large, i.e., 
the polarization of the local spin moment suppresses fluctuations. 
In the ISDW phase, Figs.~\ref{fig:scdw}(b) and~\ref{fig:scdw}(c), 
${\Delta\,{S_z^2}}$
exhibits a wave pattern consistent
with the incommensurate wave vector $q^*$ found in the spin structure factor.
The rapidity of the decay of the fluctuations away from the ends
depends on the location within the ISDW phase;
for larger ${U}$ the SDW is more pronounced, and the modulations
reach more deeply into the bulk of the system. 
Note that the charge-density fluctuations are comparably small in the
ISDW phase, albeit not as small as in the AF phase.
In the metallic phase, Fig.~\ref{fig:scdw}(d), fluctuations in the
spin density are relatively small, but still present, and fluctuations
in the charge density are significantly larger than in the other
phases, and have a high-frequency component that extends to the center
of the lattice.

\begin{figure}
	\includegraphics[width=8.6cm]{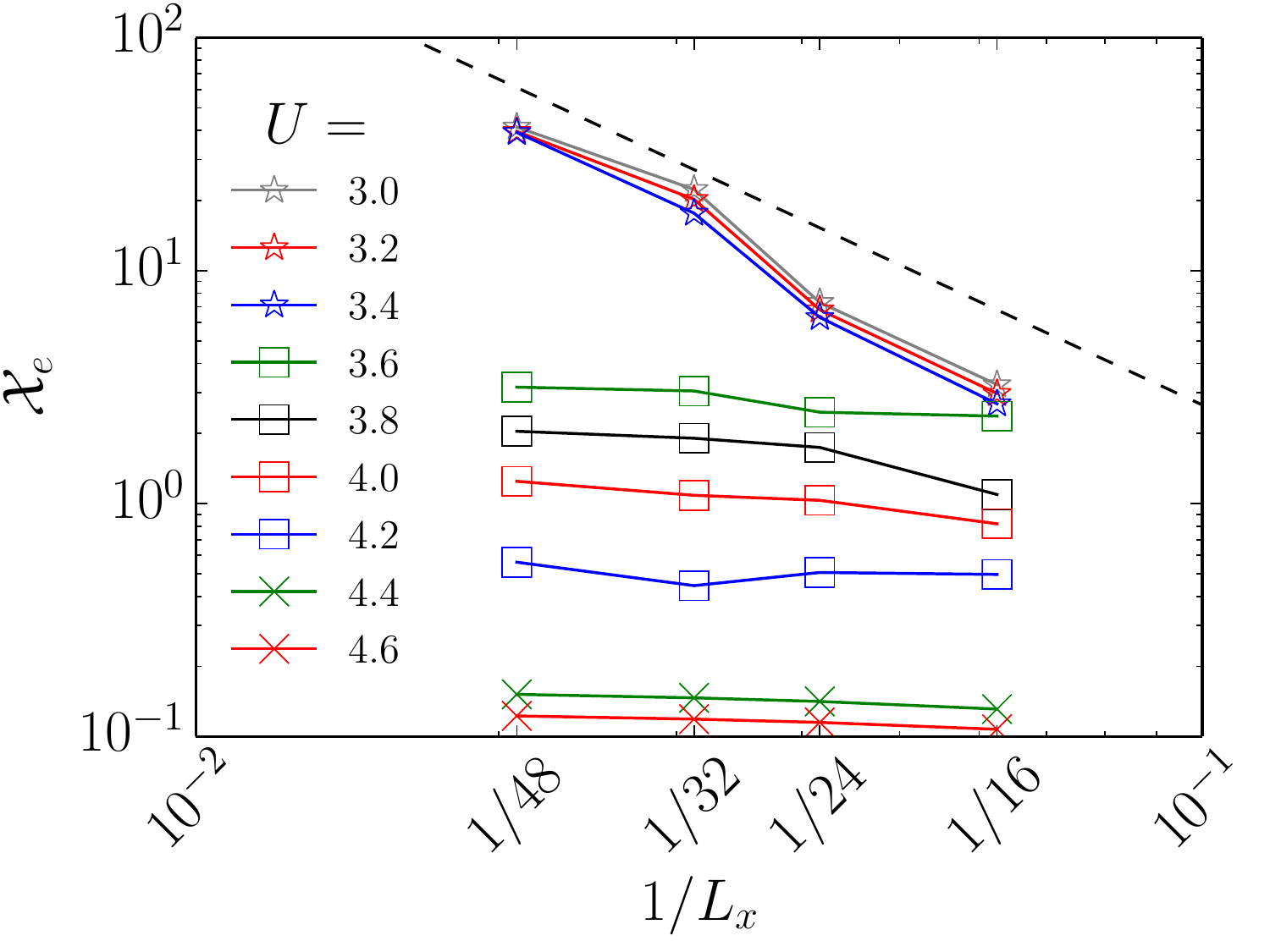}
	\caption{
		(Color online) 
		Electrical susceptibility $\mathcal{X}_e$ 
		[Eq.~\eqref{eqn:elecSusc}]
		of width-4 cylinders at ${t_{\perp}=0.7}$
		as a function of the inverse system length ${1/L_x}$
		on a log-log scale.
		The dashed line shows the analytic result for the 
		noninteracting (conducting)
		one-dimensional Hubbard model~\cite{noack2001Dielectric}.
	}
	\label{fig:elecSuscep}
\end{figure}
In order to investigate the metallic nature of all three phases, 
we have calculated the electric susceptibility $\mathcal{X}_e$ 
for different system lengths $L_x$.
By applying an electric field $E$ in the longitudinal direction, 
we can measure the polarization 
\begin{equation}
	{P=\sum_{x\,\sigma} \,
	x \, 
	\langle \Psi \, 
	| \, 
	n_{x\,\sigma} \, 
	|  \, 
	\Psi \rangle } \, ,
\end{equation}
which, in turn, allows us to obtain the electrical susceptibility
\begin{equation}
	\mathcal{X}_e
	=\frac{P}{E\,L_x} \, ,
	\label{eqn:elecSusc}
\end{equation}
assuming that $E$ is chosen small enough 
so that the response is in the linear regime.
(See Ref.~\cite{noack2001Dielectric} for a detailed description of the method.)
Figure~\ref{fig:elecSuscep} depicts the scaling of $\mathcal{X}_e$
with the inverse system length ${1/L_x}$ 
for different ${U}$ and ${t_{\perp}=0.7}$.
In the metallic phase, 
$\mathcal{X}_e$ grows proportionally to ${L_x^2}$, as can be seen from
its approximately linear behavior with a slope of $-2$ on the log-log
scale.
Such a scaling is characteristic of quasi-one-dimensional metallic
behavior~\cite{noack2001Dielectric}.
In contrast, 
in the AF phase $\mathcal{X}_e$ clearly saturates as $L_x$ becomes larger 
(i.e., ${1/L_x\to 0}$).
In the ISDW phase,
$\mathcal{X}_e$ is significantly larger
than in the AF phase, but still saturates as $L_x$ becomes larger,
thus indicating that the ISDW phase is, indeed, insulating.
Note that,
since the determination of the electrical susceptibility becomes 
progressively more difficult with decreasing interaction strength,
this analysis cannot be continued 
along the transition lines towards smaller $t_\perp$.

\begin{figure}
	\includegraphics[width=8.6cm]{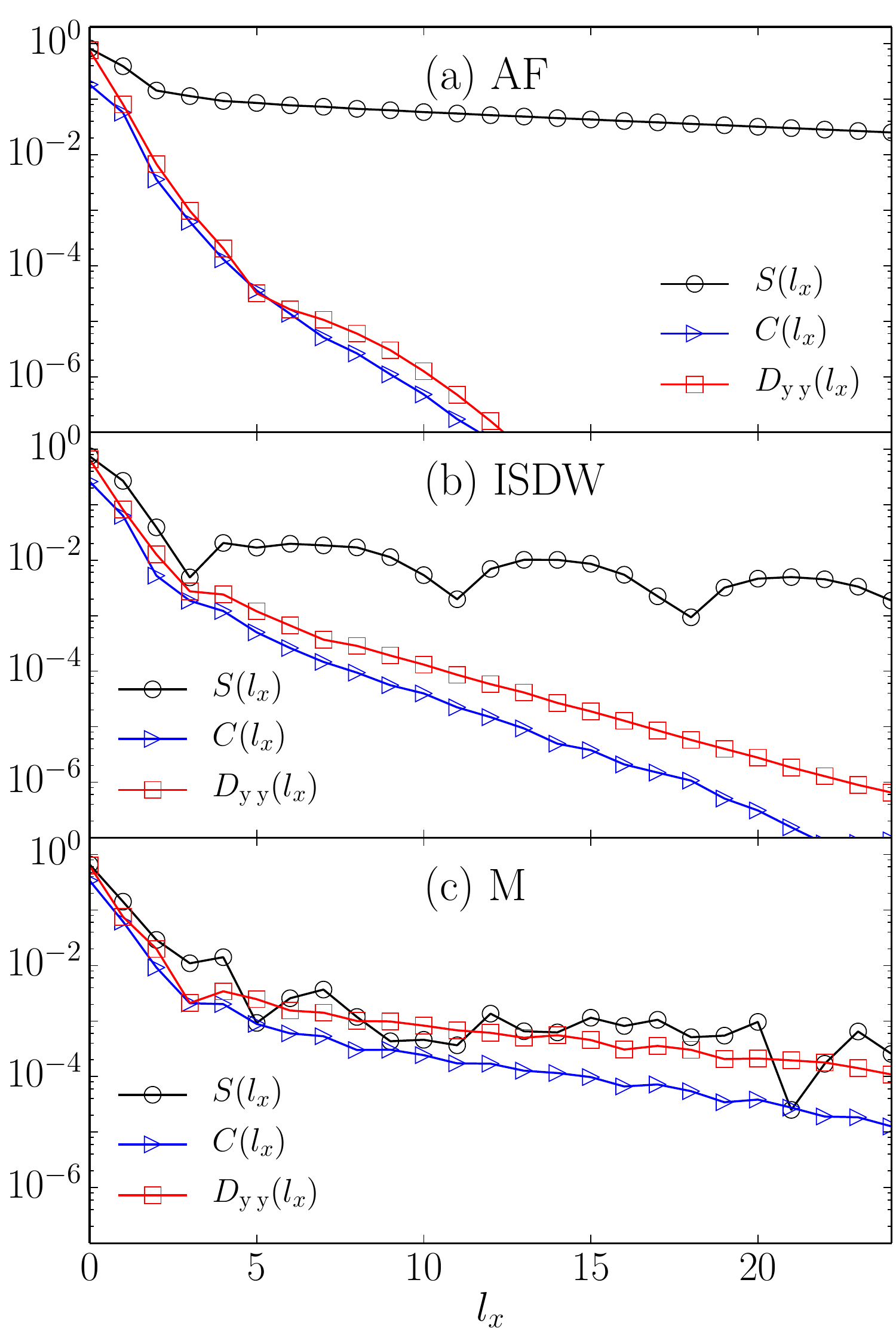}
	\caption{
		(Color online) 
		Equal-time spin, charge, and pairing correlations,
		${S(l_x)}$, ${C(l_x)}$, and ${D_{\rm y\,y}(l_x)}$ [Eqs.~\eqref{eqn:correlationfunctions}],
		as a function of the longitudinal distance $l_x$,
		for system size ${32{\times}4}$,
		transverse hopping ${t_{\perp}=0.7}$, 
		and interaction strengths
		(a) ${U=4.8}$, (b) ${U=4.0}$, and (c) ${U=3.0}$.
	}
	\label{fig:correlations}
\end{figure}
Finally, we address the question of 
whether $d$-wave pairing correlations are enhanced
in any part of the phase diagram.
Since the system is quasi-one-dimensional, we, in general,
expect at most power-law decay of the correlation functions with
distance and must compare the strength of the decay in the different
channels.
We thus calculate the equal-time spin, charge, and pair correlation functions,
\begin{align}
	S({\bf r},{\bf r}') & = 
	\langle 
	S_z({\bf r}) \,
	S_z({\bf r}') 
	\rangle 
	- \langle S_z({\bf r}) \rangle \,
	\langle   S_z({\bf r}') \rangle \, ,\nonumber \\
	C({\bf r},{\bf r}') & = 
	\langle 
	n({\bf r}) \,
	n({\bf r}') 
	\rangle 
	- \langle n({\bf r}) \rangle \,
	\langle   n({\bf r}') \rangle \, , \nonumber \\
	D_{\rm y\,y}({\bf r},{\bf r'}) & = 
	\langle 
	\Delta^{\dagger}_{\rm y} ({\bf r})  \,
	\Delta^{\vphantom{\dagger}}_{\rm y} ({\bf r'}) 
	\rangle \, ,
	\label{eqn:correlationfunctions}
\end{align}
with the local spin density, local charge density, 
and pair creation operators
\begin{align}
	S_z(x,y) & = 
	c^{\dagger}_{x\,y\,\uparrow} \,
	c^{\vphantom{\dagger}}_{x\,y\,\uparrow}
	-c^{\dagger}_{x\,y\,\downarrow} \, 
	c^{\vphantom{\dagger}}_{x\,y\,\downarrow} \, , \nonumber \\
	n(x,y) & = 
	c^{\dagger}_{x\,y\,\uparrow} \,
	c^{\vphantom{\dagger}}_{x\,y\,\uparrow}
	+c^{\dagger}_{x\,y\,\downarrow}\,
	c^{\vphantom{\dagger}}_{x\,y\,\downarrow} \, , \nonumber \\
	\Delta^{\dagger}_{\rm y} (x,y) & = 
	\frac{1}{\sqrt{2}} (
	c^{\dagger}_{x\,y+1\,\uparrow} \, 
	c^{\dagger}_{x\,y\,\downarrow}
	-c^{\dagger}_{x\,y+1\,\downarrow} \,
	c^{\dagger}_{x\,y\,\uparrow}) \, .
\end{align}
Figure~\ref{fig:correlations} shows the decay of the correlation functions 
as a function of distance in the longitudinal direction, ${l_x=|x-x^\prime|}$.
In the AF and ISDW phases,
Figs.~\ref{fig:correlations}(a) and~\ref{fig:correlations}(b),
respectively, the spin correlations 
are markedly dominant, as expected.
In the ISDW phase, the modulations of the spin correlations
are compatible with the wave pattern seen in the local spin fluctuations
in Fig.~\ref{fig:scdw}(b).
In the metallic phase, Fig.~\ref{fig:correlations}(c), 
all correlations decay at approximately the same rate.
Thus, there is no clearly dominant correlation.
Note, however, that we have not systematically investigated the
dependence of the relative strength of the correlations 
on the model parameters within the metallic phase. 

\subsection{VCA results}
\label{sec:vca_results}

\begin{figure}	
	\includegraphics[width=6.6cm]{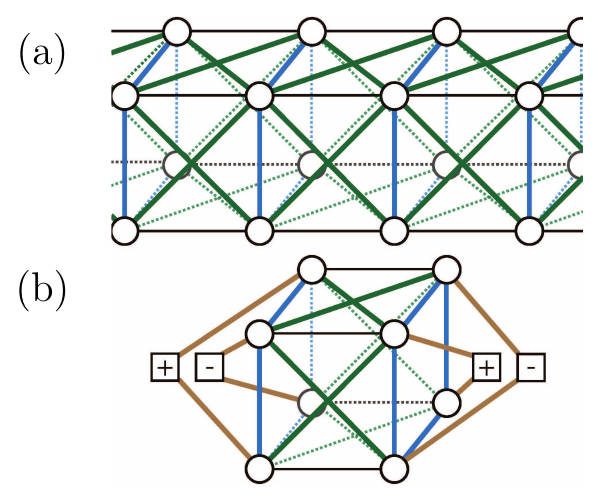}
	\caption{
	  (Color online) 
	  (a) The four-leg cylinder, showing nearest- and
      next-nearest-neighbor connections.
      (b) The ${2{\times}4}$ cluster with four bath sites, 
      periodically repeated in the $x$ direction. 
	  On the noninteracting bath sites, an antiferromagnetic Weiss
      field, indicated in the sketch by plus and minus signs, is applied.
	}
	\label{fig:VCA_cluster}
\end{figure}

In this section we apply the VCA to the Hamiltonian of Eq.~\eqref{eqn:hubbard} 
on a four-leg tube geometry, Fig.~\ref{fig:VCA_cluster}(a),
to obtain results that can be compared with 
the hybrid-space DMRG results on width-4 cylinders.
Our reference system consists of a ${2{\times}4}$ cluster 
with periodic boundary conditions in the $y$ direction, 
which is coupled to four bath sites, as depicted in 
Fig.~\ref{fig:VCA_cluster}(b).
In order to find 
the stationary point of the self-energy functional,
we use the hybridization $V$ between interacting cluster sites 
and bath sites
\footnote{
  As shown in Ref.~\cite{balzer2009}, using the hybridization $V$ 
  as a variational parameter is essential to allow for the
  coexistence of insulator and metal in a region around the
  (paramagnetic) MIT.
  The insulating and metallic solutions then differ in $V$.},
the chemical potential $\mu^{\prime}$ of the cluster
\footnote{
	Only when using $\mu^{\prime}$ as a variational parameter, 
	the determination of the electron density $n$ is thermodynamically stable: 
	The calculation of $n$ via the Green's function 
	and via a functional derivative of the self-energy functional 
	leads to the same result~\cite{aichhorn2006}.},
the chemical potential $\mu$ of the system
\footnote{
  Following the procedure introduced in Ref.~\cite{balzer2010}, 
  we Legendre transformed the self-energy functional with respect to $\mu$ 
  to be able to set a electron density to that of half filling. 
  The corresponding chemical potential $\mu$ 
  can then be determined via the variational principle.},
and the strength of an antiferromagnetic Weiss field on the bath sites
\footnote{
  We have checked that choosing a Weiss field 
  that acts on the cluster sites  instead leads 
  to qualitatively same results for the AF phase at ${t_{\perp}=1}$.
  Choosing the Weiss field to act on the bath sites 
  is in the style of cluster dynamical mean-field theory, 
  where the antiferromagnetism also enters via the bath~\cite{tremblay2006frustratedHubbard}.}
as variational parameters.

\begin{figure*}
	\includegraphics[width=18.0cm]{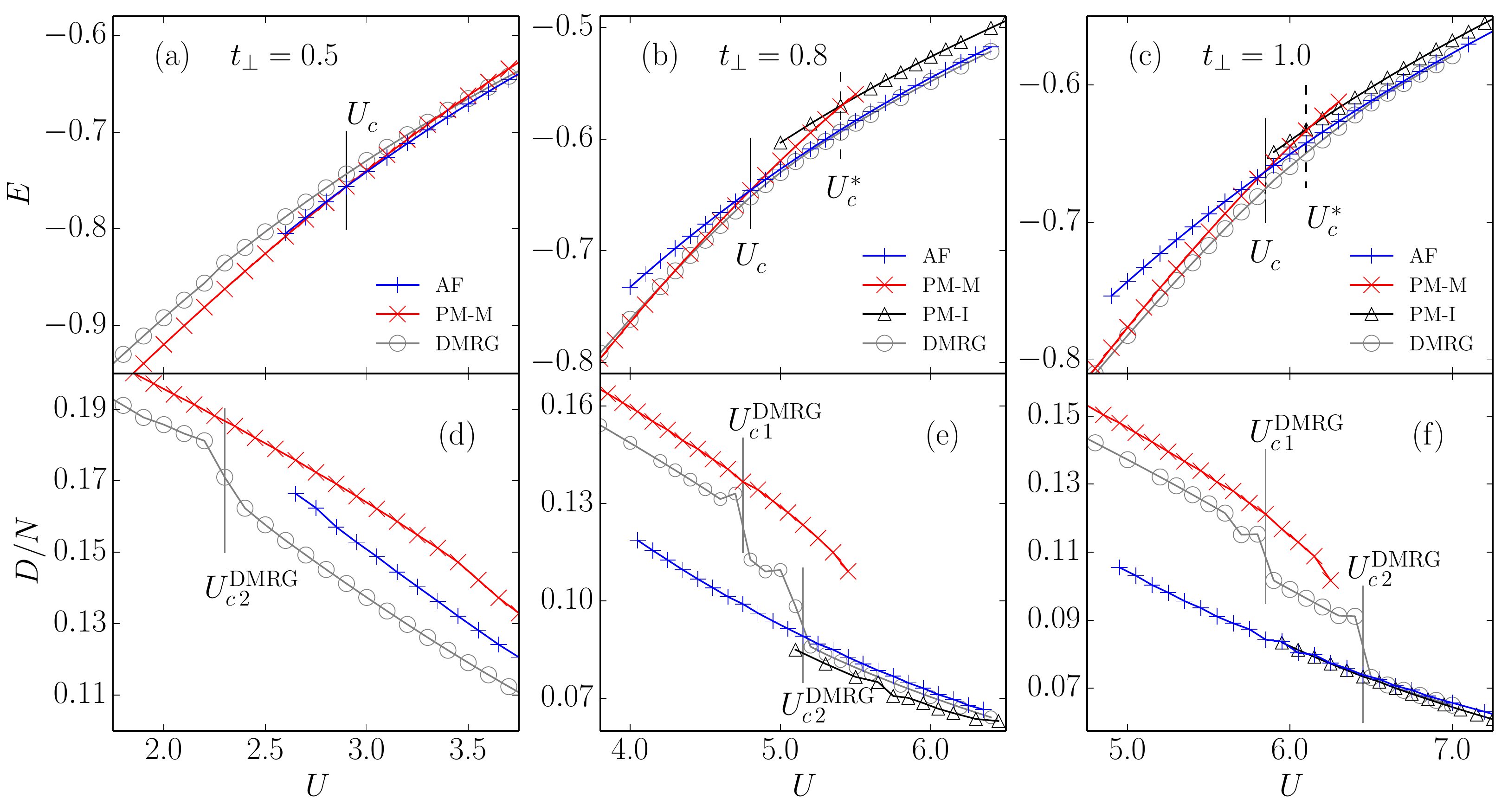}
	\caption{
		(Color online)
		Panels (a)--(c) energy density and 
		(d)--(f) double occupancy
		obtained using the VCA for the paramagnetic-metallic (PM-M), 
		paramagnetic insulating (PM-I),	and AF solutions
		for ${t_{\perp}=0.5}$, ${t_{\perp}=0.8}$, and ${t_{\perp}=1.0}$,
		as indicated.
		The black solid and dashed lines designate 
		the metal-insulator transition between the PM-M and AF
		and between the PM-M and PM-I solutions, respectively.
		DMRG results on ${16{\times}4}$ cylinders
		are plotted for comparison.
		Here, the gray vertical lines indicate the 
		transitions between the metallic and ISDW phases
		and between the ISDW and AF phases at ${U^{\rm DMRG}_{c\,1}}$
		and ${U^{\rm DMRG}_{c\,2}}$, respectively.
	}
	\label{fig:VCA_DMRG}
\end{figure*}
Figure~\ref{fig:VCA_DMRG}
depicts the energy and double occupancy
for the different solutions found by the VCA for 
interchain couplings ${t_{\perp}=0.5}$, ${t_{\perp}=0.8}$, and 
${t_{\perp}=1.0}$, together with DMRG results for ${16{\times}4}$ cylinders.
Since the VCA only captures correlations within the cluster exactly, 
the ISDW phase found in the DMRG 
cannot be seen within the VCA. 
Within the VCA, 
at interaction strengths below a critical value ${U^{\phantom{*}}_c}$, 
a paramagnetic (PM) metal is realized;
at larger interaction strengths,
the system is found to be an antiferromagnetic insulator.

A paramagnetic insulating solution is found at large interaction strengths 
(black line in Fig.~\ref{fig:VCA_DMRG}), which, however, 
is always higher in energy than the antiferromagnetic insulator 
and is therefore not realized.
Note that we find a crossing of the energies of the two paramagnetic solutions
at $U_c^*$, which is marked by the black dashed lines in 
Figs.~\ref{fig:VCA_DMRG}(b) and~\ref{fig:VCA_DMRG}(c).
The crossing is associated with a jump in the double occupancy at $U_c^*$.
This, together with the substantial coexistence region of the
paramagnetic insulating and metallic solutions around $U_c^*$, 
shows that the relatively small number of bath sites 
used in our cluster is indeed sufficient to capture 
the discontinuous transition in the paramagnetic channel.
In contrast, 
using the VCA without bath sites not only leads 
to a continuous paramagnetic metal-insulator transition 
with a much smaller $U_c^*$, but also yields a ${U^{\phantom{*}}_c}$ 
that is significantly smaller (e.g, ${U_c=3.8}$ for ${t_{\perp}=0.8}$).

Comparing the VCA data to the DMRG results, we find that,
for ${t_{\perp}=0.8}$ and ${t_{\perp}=1.0}$,
the energies of the antiferromagnetic insulator
in the VCA and in the DMRG calculation on the ${16{\times}4}$ cylinder
nearly coincide, and the agreement between the 
double occupancies in this phase is also very good.
The phase transition between the antiferromagnetic insulator
and the paramagnetic metal in the VCA occurs
within the ISDW phase seen by the DMRG,
i.e., ${U^{\rm DMRG}_{c\,1}<U^{\phantom{*}}_c<U^{\rm DMRG}_{c\,2}}$.
This is consistent to within the limitations of the methods, since, 
as discussed before, the intermediate ISDW phase cannot be found 
by the VCA for the cluster sizes treated.
Nevertheless, it is interesting to see that both 
a transition from an AF insulating phase to a PM metal 
as well as discontinuous behavior are found by both methods.
For smaller interaction strengths, the energies 
of the metallic phases within the VCA and the DMRG are again
closer to each other.
Note that the degree of numerical agreement 
has been examined here for particular (different) fixed sizes 
of the DMRG lattice and the VCA cluster;
a more complete analysis would compare results only after separate
finite-size scaling for each method.

The double occupancy of the paramagnetic metal 
is a little higher than in the DMRG.
This overestimation of the strength of the paramagnetic phase 
might be due to the fact that two cluster sites must share one bath site:
For the isotropic unfrustrated two-dimensional Hubbard model,
Balzer \textit{et al}.\ found that either the metallic or the
insulating phase is preferred, depending on whether an even or an odd
number of bath sites is coupled to each cluster site~\cite{balzer2009}.
We expect that the choice of the number of bath sites 
would have a similar effect in the present case. 
In order to check this assumption, one would have to treat
a cluster that includes at least eight bath sites.
The fact that
we are using a band-Lanczos solver at zero temperature 
and a variational space of dimension 4
means that treating the 16-site cluster would, 
unfortunately, have a much higher computational cost, 
precluding a systematic investigation of this question.

Whereas the DMRG and the VCA are in good quantitative agreement 
for the system or cluster sizes used, respectively, 
for large values of $t_\perp$,
qualitative as well as quantitative differences are present at
intermediate transverse hopping, ${t_{\perp}=0.5}$.
At relatively strong coupling, 
the VCA ground state remains an antiferromagnetic insulator, 
just as in the DMRG analysis.
A comparison of the energies of the AF solution of the VCA
and the (AF) DMRG ground-state energy down to the transition 
to a ISDW again yields good agreement.
However, while the DMRG finds the ISDW phase for moderate to low $U$,  
the VCA finds a paramagnetic metal at small $U$, 
yielding a metal-insulator transition at ${U_c\approx 2.9}$.
This transition point is comparable to the value of ${U\approx 2.3}$ 
for the transition from the AF insulator to the ISDW obtained by the DMRG.
Since the VCA on the treated cluster size is fundamentally unable 
to find an ISDW phase, the AF-to-ISDW transition cannot be seen.
Nevertheless, it is interesting to see that the VCA identifies a phase transition in a similar parameter region as the DMRG.

\begin{figure}
	\includegraphics[width=8.6cm]{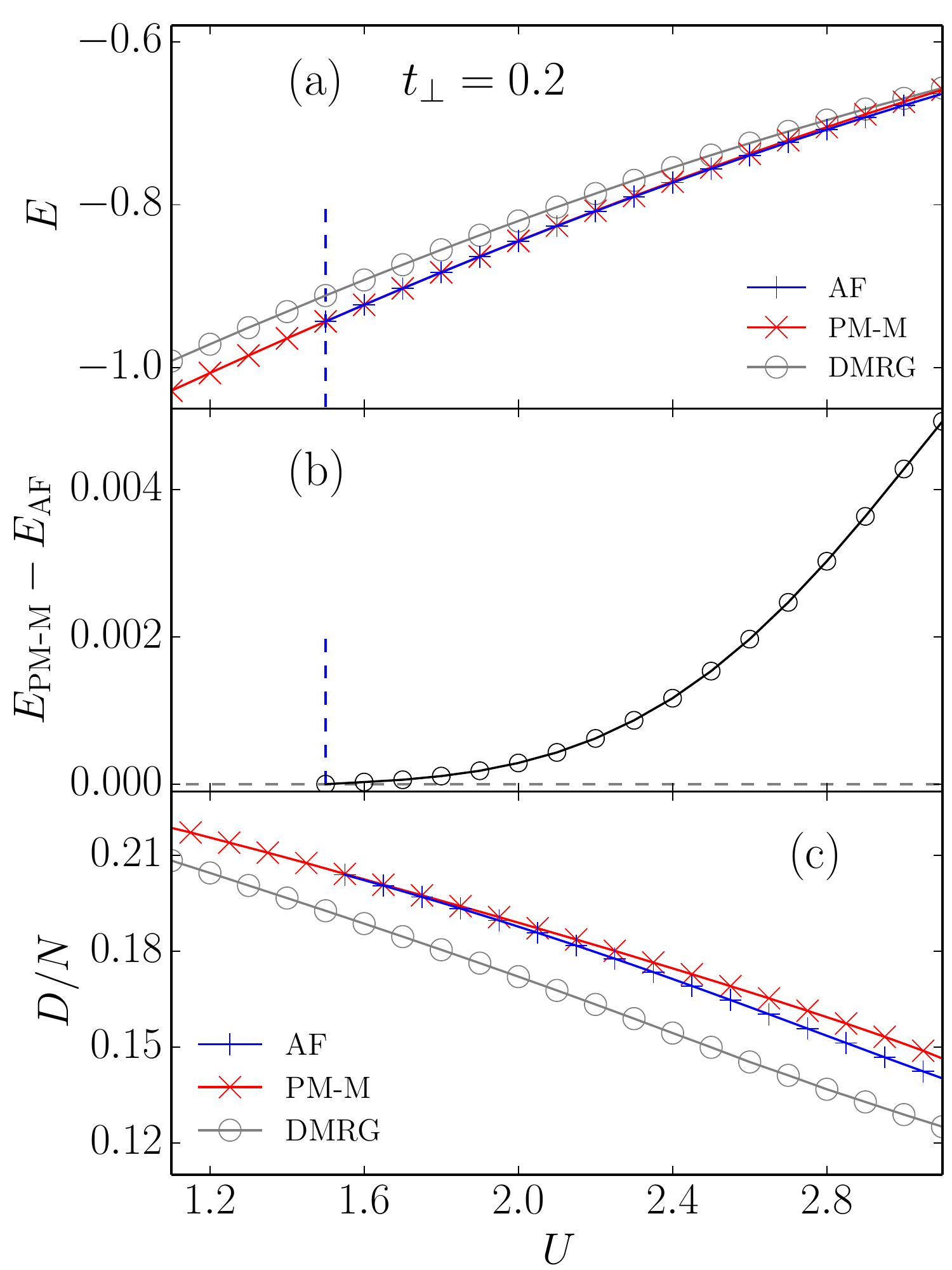}
	\caption{
		(Color online)
		(a) Energies from the VCA and the DMRG,
		(b) difference in energy between the antiferromagnetic
		and the paramagnetic solutions obtained by the VCA,
		and (c) double occupancy
		for ${t_{\perp}=0.2}$.
		The blue dashed vertical line indicates the 
		interaction strength at which the antiferromagnetic 
		and paramagnetic solutions become energetically indistinguishable 
		for the accuracy accessible within the VCA.
	}
	\label{fig:VCA_TP02}
\end{figure}
Finally, Fig.~\ref{fig:VCA_TP02} shows how the situation changes
for an even smaller value of the interchain hopping, ${t_{\perp}=0.2}$.
In the VCA, an antiferromagnetic solution is still realized for ${U>1.5}$,
Fig.~\ref{fig:VCA_TP02}(a),
but, in contrast to the previous cases at larger ${t_{\perp}}$,
the energies of the AF insulator and PM metal approach each other smoothly
as $U$ is decreased, Fig.~\ref{fig:VCA_TP02}(b),
and a level crossing cannot be observed.
At ${U=1.5}$, both solutions have almost the same energy
${(|E_{\text{PM-M}}-E_{\text{AF}}|<10^{-5})}$,
and the double occupancy exhibits no measurable jump 
at the transition point, as can be seen in Fig.~\ref{fig:VCA_TP02}(c).
For ${U<1.5}$, an AF solution cannot be found within the VCA.
For the same $t_{\perp}$, the DMRG results show
that the AF ordering in the form of a distinct peak in ${S_S({\bf k})}$ 
at ${{\bf k}=(\pi,\pi)}$ smoothly vanishes 
around the same interaction strength, ${U\approx1.5}$;
see Figs.~\ref{fig:phaseDiagram}(a) and~\ref{fig:spinPeak}.
Since, within the VCA, it is not possible to introduce a Weiss field 
that captures the quasi-one-dimensional order without introducing 
some kind of order in $y$ direction, the Q1D-AF cannot be reproduced. 
Still, the VCA finds a vanishing AF order and 
consequently a crossover to a paramagnetic phase.

\section{Discussion}
\label{sec:discussion}

We now discuss the relation between our results and those of other studies. 
For systems of a finite number of coupled Hubbard chains with varying
interchain coupling strength, a number of calculations 
using bosonization and renormalization group 
techniques in the weak-coupling limit have been carried
out~\cite{balents1996WeakChains,balents1997WeakNChains,solyom2006ladders,arrigoni1996phase}.
The case of four periodically coupled chains was
treated in detail in Ref.~\cite{balents1997WeakNChains}.
However, the half filled system, for which additional umklapp processes are
relevant, was not handled explicitly.
In addition, next-nearest-neighbor hopping, which generates
frustration and is essential to our study here,
was also not included.
Thus, we know of no weak-coupling studies that would be directly applicable to
the system studied here.
Such calculations would be extremely useful, in particular, 
in clarifying the role of the umklapp processes within the ISDW phase
and in determining the phases present in the ${U\rightarrow0}$ limit.
However, such a study would go beyond the scope of the present work.

For the isotropic case, a number of studies are available
that investigate the phase diagram of the two-dimensional Hubbard model 
as a function of the level of frustration.
For the value of the frustration treated here, ${{t'}=-0.5}$, 
VCA calculations of Nevidomskyy
\textit{et al}.~\cite{tremblay2008frustratedHubbard},
in which no bath sites were included,
found a transition from a superconducting phase to a 
${(\pi,\pi)}$ antiferromagnet,
with an intermediate region of phase coexistence 
for ${3.0\leq U \leq 4.0}$.
For the same level of frustration,
Mozusaki and Imada reported a transition 
from a metallic phase to an antiferromagnetic insulating phase
separated by an insulating, nonmagnetic, intermediate phase between
${U\approx4.0}$ and ${U\approx 6.0}$
in their PIRG study~\cite{imada2006frustratedHubbard}.
In another study using the VCA without bath sites, Yamada \textit{et al}.\ 
found either a transition between a metallic phase and 
an insulating antiferromagnetic or between a metallic phase and an
insulating paramagnetic phase~\cite{yamada2013frustratedHubbard},
depending on the size of the cluster considered.
The above studies also treated stronger levels of frustration, at which
they found different antiferromagnetic phases:
for ${t' \gtrsim 0.8}$, all three studies find
a ``striped'' phase, ${{\bf q}=(\pi,0)}$,
and the PIRG and, depending on the cluster size, the VCA, find
antiferromagnetic ordering with periodicity of two lattice sites,
${{\bf q}=(\pi,\pi/2)}$, in the region between
the ${(\pi,\pi)}$ and the ${(\pi,0)}$ phases.

In Ref.~\cite{lenz2016dimensionalCrossoverFrustratedHubbard},
the VCA with bath sites was used at zero temperature and the CDMFT at
finite temperatures to study the anisotropic model restricted to the
paramagnetic case.
These calculations yield a zero-temperature phase diagram containing
two phases: a paramagnetic metallic phase at small $U$ and a paramagnetic
insulating phase at large $U$ for all ${t_{\perp}>0.0}$.
In particular, the transition between the two phases was found to occur 
at finite values of the Coulomb repulsion for all $t_{\perp}$. 
Interestingly, the transition was found to be continuous for
${t_{\perp} \lesssim 0.2}$ and discontinuous for ${t_{\perp}\gtrsim 0.2}$.

Dimensional crossover in weakly coupled chains at finite temperature was 
studied by Raczkowski \textit{et al}.~\cite{pollet2015DimCross} using DQMC.
For ${U=2.3}$ the authors found a transition between quasi-one-dimensional 
and two-dimensional antiferromagnetic ordering similar to the one we find.

In the context of the two-dimensional Hubbard model,
it is important to address the limitations of our calculations in
understanding the discrepancies with other calculations using different methods.
The primary limitation of our calculations was that of small system width.
Indeed, we focused on systems of width 4, although, as mentioned above,
calculations for ${t_\perp=0.7}$ were also carried out for width 5.
This raises the question of what features of our phase diagram are
robust with regard to scaling in the lattice width.
In particular, it is possible that some features are remnants of the
quasi-one-dimensional nature of the lattice, i.e., of four-chain physics. 

The first feature to be considered is the presence of the intermediate 
ISDW phase.
As shown, umklapp processes are available for the width-4 cylinders
that involve a wave vector that is compatible with that of
the incommensurate  structure of the magnetic correlations 
as found in the DMRG calculations for low $U$.
This indicates that the ISDW phase 
might indeed be a feature of the four-chain system. 
On the other hand, 
width-5 cylinders for ${t_{\perp}=0.7}$ 
also show the signatures of both transitions, 
metallic phase to ISDW and ISDW to antiferromagnetic phase.

The insulating ISDW phase with incommensurate magnetic ordering 
found in our DMRG calculations has, at least up to now, not been found 
in treatments of Hamiltonian~\eqref{eqn:hubbard} in two dimensions.
It is possible that this discrepancy is due to 
the small cluster or lattice sizes used in the other calculations, 
including the VCA results presented in Sec.~\ref{sec:vca_results} of this work. 
This argument can be supported by considering other work treating the
Hubbard model on another two-dimensional, anisotropic, frustrated
lattice, namely the anisotropic triangular lattice.
Treating ${12{\times}12}$ anisotropic triangular lattices, 
Wang \textit{et al}.\ found magnetic ordering in the half filled system 
at different wave vectors away from ${(\pi,\pi)}$
in DQMC calculations~\cite{tianxing2016frustratedHubbard}.
In addition, a recent treatment using dynamical mean-field theory
after a local spin-rotating gauge transformation was applied to the system
finds a phase with incommensurate magnetic ordering for ${t'>0.7}$
for the triangular geometry~\cite{goto2016incommensurateAnisotropicHubbard}.
Incommensurate spiral magnetic ordering 
was also found by Tocchio \textit{et al.}
in a variational Monte Carlo calculation \cite{tocchio2013}. 

Furthermore, mean-field Hartree-Fock and slave-boson calculations 
find phase diagrams for the frustrated square lattice 
that feature incommensurate
phases~\cite{igoshev2010HubbardISDW,igoshev2013,igoshev2016}.
In particular, Ref.~\cite{igoshev2016} finds 
a MIT with an incommensurate intermediate phase at half filling
with a modulation of the ordering wave vector 
that is consistent with our findings
(see Fig.~1 in Ref.~\cite{igoshev2016}).

The second feature is the discontinuous nature of both transition lines,
which for the four-chain case extends over the entire length of the lines.
Here we note that the discontinuous nature of the transition 
between the AF and ISDW phases appears to be stable with regard to
scaling in the system length,
whereas it is uncertain how the transition
between the ISDW and metallic phases scales in the infinite-length limit.
However, we cannot perform a rigorous extrapolation 
of the jump in double occupancy at both transitions 
with system size and therefore cannot
completely rule out either transitions becoming continuous somewhere
along the transition lines.

The third feature is that both transition lines curve down,
reaching the ${U=0}$ axis at finite $t_{\perp}$; 
see Fig.~\ref{fig:phaseDiagram}(a).
The lower transition line, i.e., that between the metallic and the ISDW phases, 
goes to zero $U$ at a value of $t_{\perp}$ 
close to a Van Hove singularity.
Note that for the two-dimensional lattice, the Van Hove singularity
associated with a change from open to closed topology of the Fermi
surface occurs at ${t_\perp \approx 0.622}$, while 
for the four-chain system, 
the related Van Hove singularity associated with a four-band to three-band
transition occurs at ${t_\perp \approx 0.707}$.
This three-band--four-band transition point could have consequences in
a weak-coupling picture.
In particular, it is possible that umklapp processes could lead to a
Mott transition at arbitrarily weak $U$ in the four-band regime
(${t_\perp \lesssim  0.707}$), but not in the three-band regime, leading
to a Mott transition as a function of $t_\perp$ at weak $U$.

The potential behavior of the upper transition line (between the ISDW and
antiferromagnetic phases) upon system-width scaling is unclear.
Methods that are not restricted to small finite widths, in particular,
the VCA~\cite{lenz2016dimensionalCrossoverFrustratedHubbard},
obtain lines for the transition from the paramagnetic or
antiferromagnetic insulators to the paramagnetic metal that lie above
our upper transition line and do not curve down to the ${U=0}$ axis at
finite $t_\perp$, suggesting that the behavior of our transition line
could be a finite-width effect.
Since we cannot perform an extrapolation in the lattice width at that
point, we avoid further speculation about the possible outcome of such
extrapolation.
As mentioned above, weak-coupling renormalization-group
calculations could help clarify these questions in the weak-$U$ regime. 
In order to make statements at larger $U$, it would be desirable to be
able to accurately treat systems of larger width with the numerical methods;
however, significant improvement in the algorithms would be needed to do this.

\section{Conclusion}
\label{sec:conclusion}

We have investigated the ground-state phase diagram 
of the frustrated anisotropic Hubbard model 
for the case of four periodically coupled chains,
i.e., width-4 cylinders,
at half filling 
when increasing the interchain coupling
with fixed ratio of interchain hopping and frustration,
${t'/t_\perp = -0.5}$.

Using the hybrid--real-momentum-space DMRG, 
we have mapped out the phase diagram as a 
function of the interchain hopping and the interaction strength.
In the region of weak on-site interaction and strong interchain coupling,
we find a metallic phase, 
which vanishes at the point where the noninteracting Fermi surface
undergoes a topological change at the Van Hove singularity.
At higher interaction strengths, we find an antiferromagnetic insulator
that smoothly changes from quasi-one-dimensional antiferromagnetic ordering
for weak interchain coupling to two-dimensional ordering for stronger
interchain coupling.
The quasi-one-dimensional antiferromagnetic behavior remains present 
as the interaction strength becomes weaker and even extends 
to larger interchain coupling at sufficiently weak interaction strength.
The metallic and antiferromagnetic phases are separated by an
intermediate incommensurate spin-density-wave state, i.e., 
a state characterized by dominant incommensurate spin correlations and
insulating behavior.
For fixed lattice size,
we find that the transitions between these phases are discontinuous,
as characterized by finite jumps in the double occupancy 
and the von Neumann entropy 
and discontinuous changes in the static spin structure factor.

Our numerical calculations were carried out on a lattice with
cylindrical topology and cylinders 
of lengths ranging from 16 to 48
lattice spacing and, primarily, of width 4.
Here, the discontinuous nature of the transition between the antiferromagnetic
and the incommensurate spin-density wave is stable as a function of the length,
while we cannot determine whether the transition between
the latter and the metallic phase becomes continuous in 
the infinite-length limit.
We have been able to carry out a minimal check of robustness of the
results on lattice width by comparing results for width-4 and width-5
lattices at one value of $t_\perp$, ${t_\perp=0.7}$ as a function of
interaction strength for one length, 16; the signatures of the two
transition points found on the width-4 lattices are also
present in the width-5 results.

For selected values of the interchain coupling,
we apply the VCA within a cluster geometry that corresponds to 
the cylindrical lattice geometry used in the DMRG calculations 
so that the VCA results can be compared to the unbiased DMRG results.
For intermediate to strong interchain coupling,
the VCA obtains a discontinuous phase transition
between a metallic conducting and an antiferromagnetic insulating phase
with energies and double occupancies that agree well with the DMRG results.
However, the restricted cluster sizes in the VCA 
hinders the treatment of incommensurate spin
correlations, and the VCA results hence cannot pick up
the intermediate incommensurate spin-density-wave phase. 
For weaker interchain coupling, the metal-insulator transition
becomes continuous in the VCA and occurs around the same
value of the interaction strength
at which the antiferromagnetic order with 
distinct ordering wave vector ${(\pi,\pi)}$
vanishes smoothly in the DMRG.
Thus, we find good agreement between the VCA and DMRG results in the
region where we expect it, at intermediate to strong interchain
hopping and moderate to strong interaction strength.
The absence of the incommensurate spin-density-wave-phase in the VCA
and the discrepancies at weak interchain coupling
are understood by the limitations of the VCA clusters used.
One central question raised by our study is whether the
intermediate incommensurate spin-density-wave phase is pertinent to
the two-dimensional system or, indeed, 
to finite-width cylinders of larger width.
On the one hand, we have found that the magnetic structure 
in this phase is compatible with 
possibly relevant umklapp scattering processes in width-4 Hubbard cylinders,
indicating that the intermediate phase could be specific to four-chain physics.
On the other hand, we have found some indications of the presence of
an intermediate incommensurate phase in the width-5 cylinder and, as
described above, some studies on related
two-dimensional systems also find similar intermediate phases.
Further investigation of this issue, for example, using the
renormalization group and bosonization at weak coupling, or by
jointly applying and comparing various numeric methods,
as has recently proven to be fruitful for the doped two-dimensional Hubbard
model~\cite{leblanc2015HubbardBenchmark,zheng2017stripes}, would
certainly be very interesting.

\FloatBarrier
\begin{acknowledgments}

We thank M.~Raczkowski and F.~F.~Assaad for helpful discussions,
and we acknowledge computer support by the GWDG
and the GoeGrid project.
This work was supported in part by the Deutsche
Forschungsgemeinschaft (DFG) through 
Research Unit FOR 1807, projects P2 and P7,
and the European Research Council (Project No.~617196).

\end{acknowledgments}


%

\end{document}